\documentclass[iop,apj,twocolumn]{emulateapj}
\usepackage{amsmath,amssymb,amstext}
\usepackage{aas_macros}
\usepackage{natbib}

\usepackage[breaklinks,colorlinks,citecolor=blue,linkcolor=magenta,urlcolor=blue]{hyperref} 

\usepackage[all]{hypcap} 

\usepackage{graphicx}
\usepackage{grffile}
\usepackage{soul}

\begin{document}

\title{Three-dimensional magnetohydrodynamical simulations of the morphology of head-tail radio galaxies based on magnetic tower jet model}

\author{Zhaoming Gan$^1$, Hui Li$^2$, Shengtai Li$^2$, and Feng Yuan$^1$}
\affil{$^1$ Key Laboratory for Research in Galaxies and Cosmology, Shanghai Astronomical Observatory, Chinese Academy of Sciences, \\80 Nandan Road, Shanghai 200030, China; \href{mailto:zmgan@shao.ac.cn}{zmgan@shao.ac.cn}, \href{mailto:fyuan@shao.ac.cn}{fyuan@shao.ac.cn}
\\$^2$ Theoretical Division, Los Alamos National Laboratory, Los Alamos, NM 87545, USA; \href{mailto:hli@lanl.gov}{hli@lanl.gov},  \href{mailto:sli@lanl.gov}{sli@lanl.gov}}

\shorttitle{magnetic tower jet under intra-cluster weather}
\shortauthors{Gan et al.}

\begin{abstract}
The distinctive morphology of head-tail radio galaxies reveals strong interactions between the radio jets and their intra-cluster environment, the general consensus on the morphology origin of head-tail sources is that radio jets are bent by violent intra-cluster weather. We demonstrate in this paper that such strong interactions provide a great opportunity to study the jet properties and also the dynamics of intra-cluster medium (ICM).  
By three-dimensional magnetohydrodynamical simulations, we analyse the detailed bending process of a magnetically dominated jet, based on the magnetic tower jet model. We use stratified atmospheres modulated by wind/shock to mimic the violent intra-cluster weather. 
Core sloshing is found to be inevitable during the wind-cluster core interaction, which induces significant shear motion and could finally drive ICM turbulence around the jet, making it difficult for jet to survive. We perform detailed comparison between the behaviour of pure hydrodynamical jets and magnetic tower jet, and find that the jet-lobe morphology could not survive against the violent disruption in all of our pure hydrodynamical jet models. On the other hand, the head-tail morphology is well reproduced by using a magnetic tower jet model bent by wind, in which hydrodynamical instabilities are naturally suppressed and the jet could always keep its integrity under the protection of its internal magnetic fields. Finally, we also check the possibility for jet bending by shock only. We find that shock could not bend jet significantly, so could not be expected to explain the observed long tails in head-tail radio galaxies. 
\end{abstract}

\keywords{galaxies: active --- galaxies: jet --- magnetohydrodynamics (MHD) --- methods: numerical}
\maketitle

\section{Introduction}

It is now widely believed that the energy and momentum output from active galactic nuclei (AGNs), in forms of radiation, wind and jet, plays a crucial role in the formation and evolution processes of their host galaxies \citep[a.k.a AGN feedback, see, e.g.,][]{ostriker_momentum_2010, fabian_observational_2012, gan_active_2014,king_black_2016}. Much attention is paid on the energy budget of the powerful jets \citep[up to $10^{61}$ erg, e.g.,][]{mcnamara_heating_2005, fabian_observational_2012}, which is potentially important to the dynamics of their surrounding intra-cluster medium (ICM), e.g., quenching cooling flow \citep{fabian_cooling_1994, omma_heating_2004}. However, the jet dynamics itself and the detailed jet-feedback mechanism are still not clearly understood so far, with questions regarding the jet composition, the role of magnetic fields in the jet dynamics, dissipation of jet energy into the surrounding medium, and so on.

Much effort has been made to understand the jet physics on two fundamental aspects: 
(1) the basic jet dynamics. From the very central engine \citep[black hole accretion disk, e.g.,][]{mckinney_disc-jet_2007, Yuan_numerical_2015, Tchekhovskoy_three-dimensional_2016} to large radio lobes \citep[e.g.,][]{reynolds_hydrodynamics_2002, lynden-bell_why_2003, li_modeling_2006, oneill_three-dimensional_2010}, several jet models have been proposed so far, while most models have used magnetic fields to launch jet from the central engine, they differ on energy composition on large scales;
(2) the interaction between the jet and its ambient environment (including jet feedback). It is well accepted that the ambient environment is of crucial importance to the jet propagation, especially on the hundred-kilo parsec scale where jet already slows down and forms radio lobes \citep[e.g.,][]{Heinz_answer_2006}. A special type of radio galaxies ---  head-tail radio sources --- deserves special attention, as their distinctive morphology provides a great opportunity to study the underlying physics of both the jet dynamics and its ambient environment. 

\begin{figure*}[t]
\label{fig:cartoon-head-tail-sources}
\centering
\includegraphics[height=0.375\textwidth]{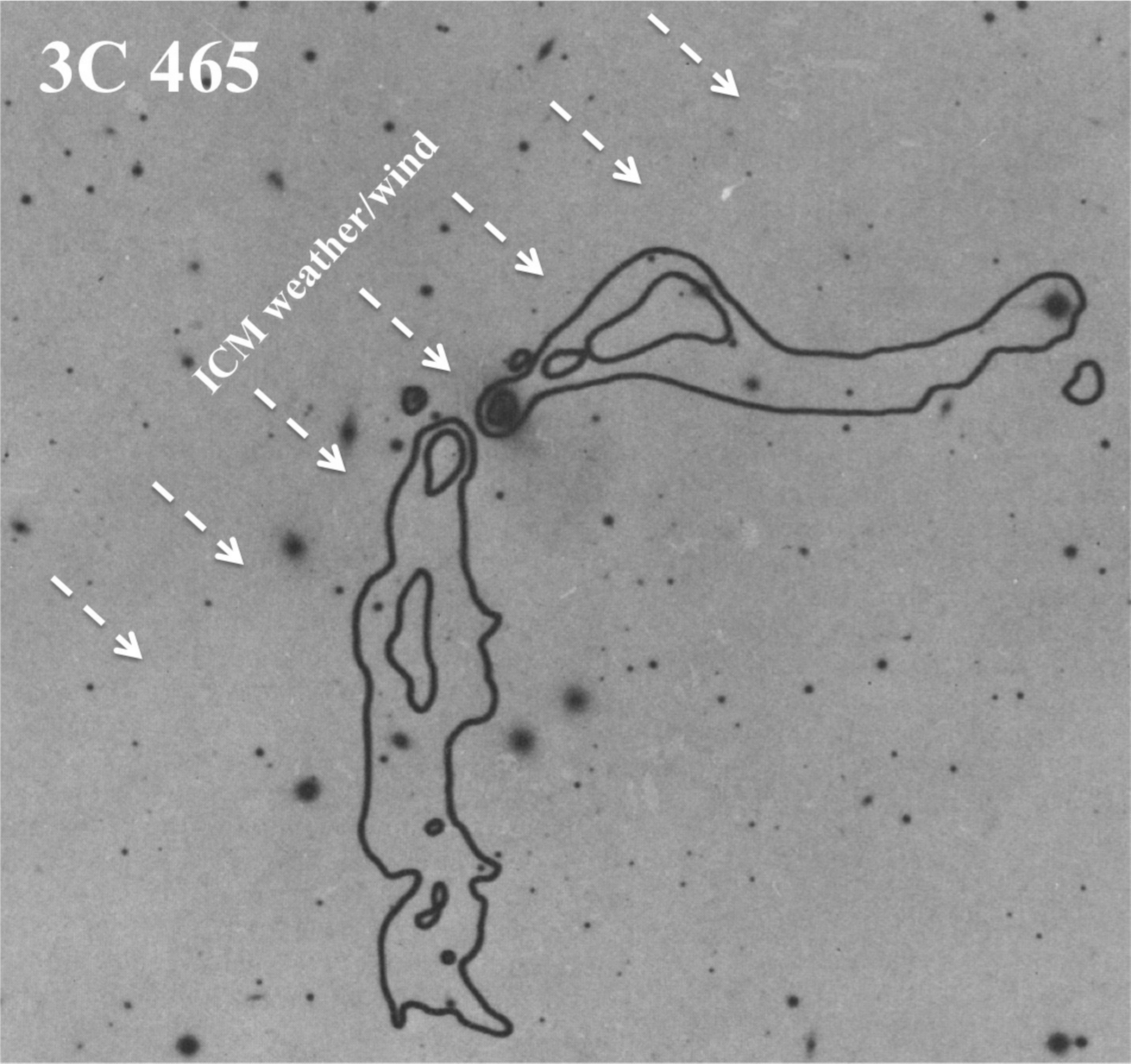}
\includegraphics[height=0.375\textwidth]{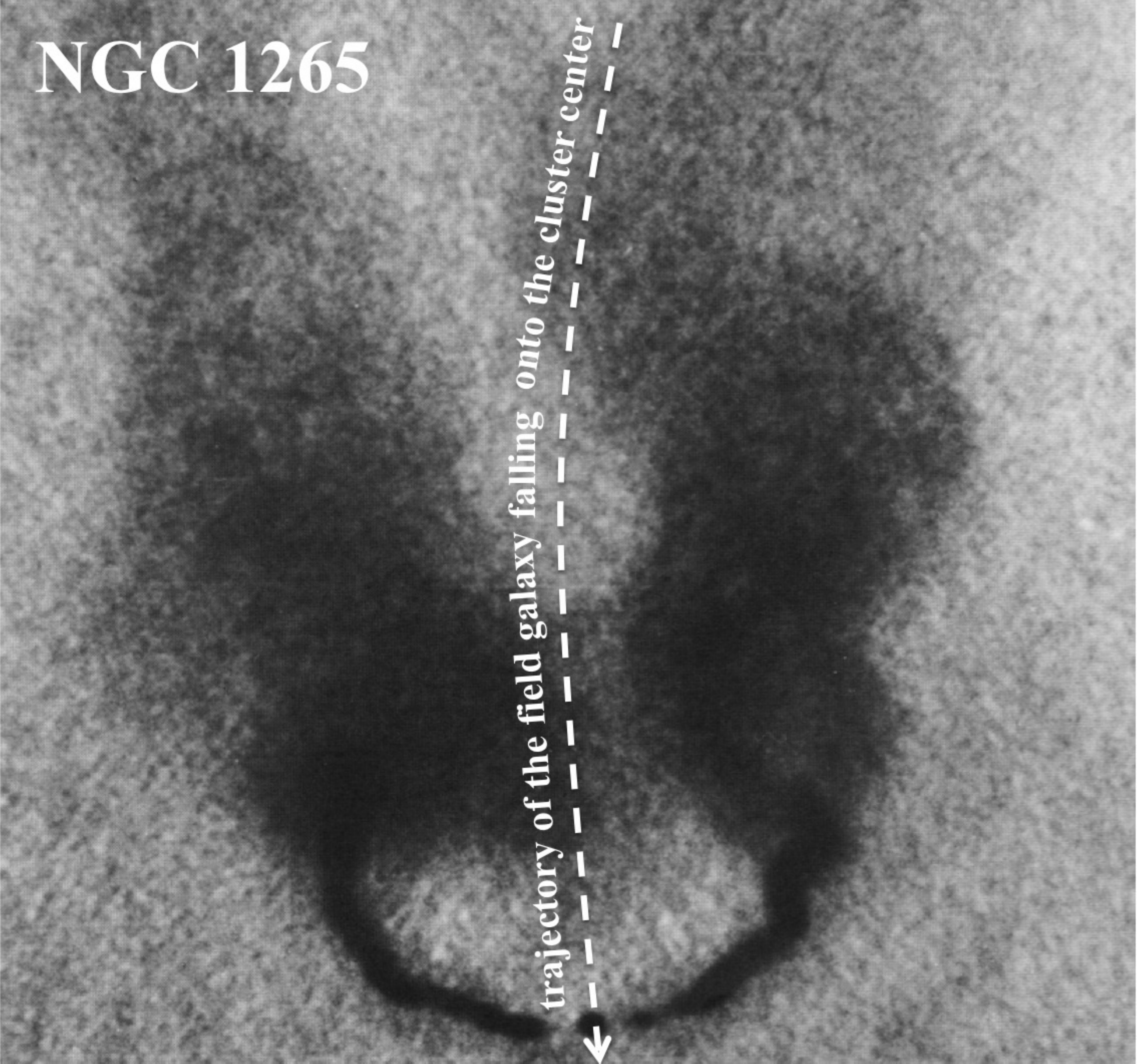}
\caption{Cartoons for jet bending in head-tail radio galaxies. Left panel: the prototype of wide-angle tailed sources, 3C 465,  where some ICM weather/wind is expected to bend the jet. The black solid lines are contour plot of radio intensity (taken from \citealp{eilek_what_1984}); Right panel: the prototype of narrow-angle tailed sources, NGC 1265, a field galaxy falling onto the center of the Perseus cluster. The image shows the white and black radiophotograph (taken from \citealp{odea_multifrequency_1986}).}
\end{figure*}

The general consensus on the morphology origin of head-tail sources is that radio jets are bent by violent intra-cluster weather \citep[see Figure \ref{fig:cartoon-head-tail-sources};][]{begelman_twin-jet_1979, jones_hot_1979, odea_multifrequency_1986}. The observed head-tail radio sources are classified into two types according to their tail morphology, i.e., the narrow-angle tailed sources \citep[NATs; NGC 1265 is known as its prototype, see Figure \ref{fig:cartoon-head-tail-sources}, right panel;][]{ryle_radio_1968, odea_multifrequency_1986, sun_small_2005} and the wide-angle tailed sources \citep[WATs; 3C 465 is known as its prototype, see Figure \ref{fig:cartoon-head-tail-sources}, left panel;][]{burns_structure_1981, eilek_what_1984, eilek_magnetic_2002, hardcastle_chandra_2005}. As the tails are bent into the same direction, it implies that the intra-cluster weather should be some regular ICM motion relative (probably transverse) to the jet, rather than turbulence.  The origin of such intra-cluster weather is subtle and there are several possibilities:  
(1) it is plausible that NATs are usually field galaxies moving through their host cluster, e.g., NGC 1265, it is falling onto the cluster center with a speed up to $\sim$ 2500 km/s \citep{sun_small_2005}; 
(2) WATs usually settle in the cluster center, i.e., at the bottom of the potential well, so their relative velocities to the background are thought to be low, e.g., the one of 3C 465 is supposed to be only few hundreds of km/s \citep{leahy_3c_1984}; 
(3) it is also known that merger between two clusters could induce shocks, the shock passage through the radio galaxies could also affect the jet morphology.

It is realised that the behaviour of different jet models under such violent intra-cluster weather could be very different.  So it could provide strong references on the jet dynamics and the ICM properties \citep[see, e.g.,][]{freeland_bent-double_2008, morsony_simulations_2013}. It is our aim in this paper to study the detailed dynamics of the jet bending process in a realistic intra-cluster environment with violent weather, using the powerful tool of massive numerical simulation.

There are so far only few numerical simulations on the head-tail sources in the literature. A light jet and supersonic/transonic wind are usually assumed in these numerical models. For example, \citet{soker_numerical_1988} used a three-dimensional  particle-in-cell scheme to simulate the bending of NATs by ram pressure or pressure gradients. They found that the pressure gradients are much less effective in bending jets when compared to ram pressure. \citet{balsara_three-dimensional_1992} simulated the bending process of a hydrodynamical jet by supersonic crosswind. They found that the Rayleigh-Taylor and Kelvin-Helmholtz instabilities could help to form the relaxed radio tails of NATs before the jets are entirely disrupted. \citet{loken_radio_1995} investigated the jet bending processes (by galactic wind/shock) of WATs in merging cluster environments. They also used a hydrodynamical jet model, and proposed that WATs could be used to mark clusters which have recently undergone a merger. Despite of the successes claimed by the authors, a problem of the hydrodynamical jet models is that jet material could spread all the way in the downstream of the ICM winds \citep[see also][]{morsony_simulations_2013}, this is not favoured by the radio observations (radio emission is hardly detected, if any, between the tails. See, e.g., Figure \ref{fig:cartoon-head-tail-sources}, the radio images of 3C 465, NGC 1265). Jones and his colleagues extended their numerical models into the MHD regime (e.g., \citealp{porter_narrow_2009} on NATs; \citealp{mendygral_mhd_2012} on WATs). However, kinetic energy always dominates over the magnetic energy in their models, which is more like a hydrodynamical jet even though there are magnetic fields in the jet. To the best of our knowledge, so far there isn't a numerical model for head-tail sources in which the jet is magnetically driven and the jet energy is dominated by magnetic energy. The latter is what we shall focus on in this paper.

A magnetically dominated jet has several distinctive features when compared to hydrodynamical jets. For example, Li and his colleagues proposed an approach to model the large-scale behaviour of jets in the magnetically dominated regime, which is also called the magnetic tower model (\citealp{li_modeling_2006, nakamura_structure_2006, nakamura_stability_2007, nakamura_numerical_2008}; see also, \citealp{ustyugova_poynting_2000, lynden-bell_magnetic_1996, lynden-bell_why_2003}). In their models, the jet is driven magnetically, and the magnetic fields are of helix-like structures (similar magnetic field configuration is also found in numerical simulations on the jet engine, i.e., black hole accretion disk, see \citealp{Yuan_hot_2014} for a review). Compared to hydrodynamical jets, magnetic fields in magnetic tower jet can help to stabilise the jet against instabilities. Under the protection of the return current,  the jet could always keep its integrity and is expected to have a relatively sharp edge in its radio image. 

When simulating the jet propagation in galactic environment, \citet{nakamura_structure_2006} found that the magnetic tower jet could naturally inflates into larger lobes when it penetrates out off the cluster core, and the lobe size is determined by the balance between the magnetic pressure within the lobes and the thermal pressure of ambient ICM. \citet{nakamura_stability_2007} analysed the detailed dynamics and instabilities during the lobe formation of magnetic tower jet, and found a tight relation between the locations of the lobes (or bubbles) and the ICM profile of theirs host clusters. \citet{diehl_constraining_2008} tested the scenario in terms of interpreting the morphology of 64 X-ray cavities in clusters and galaxies (i.e., the correlation between the X-ray bubble size and its location). They found that the magnetic tower jet model works better when compared to the hydrodynamical jet models. \citet{xu_formation_2008,xu_turbulence_2009} studied the formation of X-ray cavities in a realistic cluster environment with ICM turbulence, using cosmological MHD simulations with the magnetic tower jet model. They found that up to 80\%-90\% of the jet energy goes into doing working against the hot ICM, which is potentially very important to the formation of the host galaxy and also to the origin of the cluster magnetic fields.

In this paper, we focus mainly on the detailed jet bending process, by performing three-dimensional (3D) MHD simulations of magnetic tower jet \citep{li_modeling_2006} in realistic intra-cluster environment with violent weather. We also compare the simulation results with those of hydrodynamical jet models, in this way, we shall demonstrate that the jet-lobe morphology could be used to study the jet properties and the ICM dynamics. Without loss of generality, two kinds of intra-cluster weather are studied in this paper: (1) wind (relative motion between the radio-jet host galaxy and its surrounding ICM, like the case of NGC 1265) and (2) shock  (e.g. those induced by galaxy/cluster merger). We build cartesian coordinates co-moving with the jet central engine (i.e., the supermassive black hole), and use the King's beta model \citep{king_structure_1962} to mimic the realistic surrounding ICM atmosphere. Finally, the wind/shock is put into the computational domain through the boundary.

The plan of this paper is as follows. In \S\ref{sec:model-setup}, we briefly introduce the basic equations of the magnetic tower jet, the model setup of the galactic environment and also the intra-cluster weather. The numerical results are presented in \S\ref{sec:results}. 
In \S\ref{sec:conclusion}, we present the conclusions and also discuss some possible observational effects of our model.

\begin{table*}[ht]
\caption{Units of physical quantities for normalisation}\label{tab:units}
\begin{center}
\begin{tabular}{llll}
\hline\hline
{Physical Quantities} & {Description} &{Normalisation Units} & {Typical Values}\\
\hline
$R\,(= \sqrt{x^2+y^2+z^2})$\dotfill & Length         & $R_{0}$                   & 5 kpc                           \\ 
$\rho$             \dotfill & Density          & $\rho_{0}$                & $5.0 \times 10^{-27}$ g/cm$^3$  \\
$p$                 \dotfill & Pressure        & $p_0$                      & $3.1 \times 10^{-11}$ dyn/cm$^2$\\
${\bold v}$      \dotfill & Velocity          & $C_{\rm s0}=\sqrt{p_0/\rho_0}$   & $7.9 \times 10^7$ cm/s \\
$t$                  \dotfill & Time               & $t_0=R_{\rm 0}/C_{\rm s0}$        & $6.2 \times 10^6$ yr      \\
${\bold B}$     \dotfill & Magnetic field & $B_0=\sqrt{4\pi\rho_{0}C^{2}_{\rm s0}}$ & 19.7 $\mu$G     \\
\hline
\end{tabular}
\end{center}
All the physical quantities are normalized by setting $\rho_0=p_0=R_0=1$, where $\rho_0$ is the typical density in the system we simulate, and $p_0$ is chosen according to an effective temperature of $k_{B}T=4$ keV (with mean molecular weight $\mu=0.62$). We therefore have the unit of velocity $v_0=C_{\rm s0} = \sqrt{p_0/\rho_0} \simeq 7.9 \times 10^7 cm/s$. $R_0$ is the typical length scale of the system, which is set to 5 kpc, therefore the time unit is $t_{0} = R_{0}/C_{\rm s0} \simeq 6.2$ Myr. Finally, the unit of magnetic field is chosen so that the  magnetic permeability is unity.
\end{table*}

\section{Numerical Models} \label{sec:model-setup}

\subsection{Basic Equations} \label{sec:basic-equations}
We solve the time-dependent ideal MHD equations numerically in a three-dimensional Cartesian coordinate system ($x,y,z$) using the code LACOMPASS developed by \citet{li_alamos_2003} \citep[see also][]{li_modeling_2006, nakamura_structure_2006},
\begin{equation}
\label{eq:mass}
\frac{\partial\rho}{\partial t} + \nabla\cdot(\rho{\bf v}) = \dot{\rho}_{\rm inj},
\end{equation}
\begin{equation}
\label{eq:momentum}
\frac{\partial (\rho {\bf v})}{\partial t} + \nabla\cdot \left( \rho {\bf v v} + p + B^2/2 - {\bf B B}\right)  
= - \rho \nabla\psi + \dot{\rho}_{\rm inj} \cdot v_{\rm inj,z},
\end{equation}
\begin{equation}
\label{eq:energy}
\frac{\partial E}{\partial t} + \nabla\cdot\left[\left(E +p + B^2/2 \right){\bf v} 
-{\bf B}({\bf v}\cdot{\bf B})\right] = -\rho {\bf v} \cdot \nabla \psi + \dot{E}_{\rm inj},
\end{equation}
\begin{equation}
\label{eq:induction}
\frac{\partial {\bf B}}{\partial t} - \nabla\times( {\bf v} \times {\bf B}) = {\dot {\bf B}}_{\rm inj}.
\end{equation}
\noindent
where all variables have their usual meaning. To close the equations, we use an ideal gas equation of state with adiabatic index $\gamma = 5/3$. $\dot E_{\rm inj} = \dot{E}_{\rm inj,kin} + \dot{E}_{\rm inj,th} + \dot{E}_{\rm inj,mag}$ is the jet power in forms of kinetic energy, thermal energy and magnetic energy, respectively. 
The units and typical values of physical quantities for normalisation are listed in Table \ref{tab:units}.

\subsection{Jet} \label{sec:jet-model}
The magnetic tower jet is implemented by injecting magnetic fields into the galaxy center (Equation \ref{eq:induction}). Following \citet{li_modeling_2006}, the injection term $\dot{\bf B}_{{\rm inj}} = \gamma_b {\bf B}_{{\rm inj}}$ in the induction equation is to incorporate the external injection of magnetic fields, where $\gamma_b(t)$ specifies the time-dependent injection rate. For simplicity, we set $\gamma_b$ to be constant during the whole period of jet activity $\Delta t_{\rm inj}$. In the magnetic tower jet model, ${\bf B}_{{\rm inj}}$ is given by the equations below in cylindrical coordinate system ($z,r,\phi$),
\begin{eqnarray}
B_{{\rm inj},r}     &=& B_0 \cdot \frac{2 z r}{R_0^2} \exp\left(-\frac{r^2 + z^2}{R_0^2}\right)~~,\label{eq:br}\\
B_{{\rm inj},z}    &=& B_0 \cdot 2 (1-r^2/R_0^2)      \exp\left(-\frac{r^2 + z^2}{R_0^2}\right)~~,\label{eq:bz}\\
B_{{\rm inj},\phi}&=& B_0 \cdot \frac{\alpha r}{R_0}\exp\left(-\frac{r^2 + z^2}{R_0^2}\right)~~.\label{eq:bt}
\end{eqnarray}
The injection is confined into a limited volume as the flux drops exponentially as a function of distance from the center, and $R_0$ is the characteristic injection radius, which we also use to normalize length of our simulation domain (Table \ref{tab:units}). Parameter $\alpha$ determines the ratio between the toroidal and poloidal components of the injected magnetic fields. $B_0$ is the typical value of magnetic intensity for normalisation in our simulations (Table \ref{tab:units}).

\textcolor{black}{The injected magnetic field given by equations above is non-fore-free \citep{li_modeling_2006}. As a result, most of the gas in the injection zone will be driven out inevitably, causing the density there to decrease rapidly. To mimic the mass injection from the central engine, and also to} avoid extremely low density regions after the magnetic fields have undergone large expansion, we also inject some mass along with the magnetic fields,
\begin{equation}
\label{eq:rhoinj}
\dot{\rho}_{\rm inj} = \gamma_\rho \rho_0 \cdot \exp\left[-\frac{(r^2 + z^2)}{R_0^2}\right],
\end{equation} 
where $\gamma_\rho$ is the characteristic mass injection rate. $\rho_0$ is the typical value of the ICM density, which we use to normalize density in our simulations (Table \ref{tab:units}). The injected mass also has temperature $T_{\rm inj}$ and vertical velocity $| v_{\rm inj,z} |$. 

Practically, we control the power and energy budget of the jet by adjusting parameters $\gamma_b$, $\gamma_\rho$, $T_{\rm inj}$ and $| v_{\rm inj,z} |$, making sure that the values of jet power, total energy and final magnetic intensity lie in a reasonable range as of the observed ones.
By changing these parameters, it is possible for us to make runs for different types of jet, e.g., it degenerates to hydrodynamical jet model when $\gamma_b=0$. We could also choose proper values for $T_{\rm inj}$ and $| v_{\rm inj,z} |$ to make runs for supersonic jet and subsonic jet.  While for runs of magnetic tower jet, we always make sure that the injected magnetic energy dominates over the kinetic energy and thermal energy. 

Finally, the duration of jet injection $\Delta t_{\rm inj}$ is determined by the lifetime of AGN activity, of which the typical value is $\Delta t_{\rm inj}\sim30$ Myr.

\begin{table*}[ht]
\caption{Model parameters in the simulations.}
\label{tab:runs}
\begin{center}
\begin{tabular}{lccccccclllc}
\hline\hline
{Model} & {Wind Speed} &{$\alpha$}&{$\gamma_B$} & {$\gamma_\rho$}& $T_{\rm inj}$&$|v_{\rm inj,z}|$ &{$\Delta t_{\rm inj}$}& \multicolumn{3}{c}{Injected Jet Energy (erg)}&{Jet Power}\\ \cline{9-11} 
&(km/s)&&&&(K)&(km/s)&(Myr)&{Kinetic}&{Thermal}&{Magnetic}&(erg/s)\\ \hline
C1 & -- & 15.0 & 3.0 & 0.10 &$4.6\times10^{7}$&0& 31.0 &  0 & $1.02\times10^{56}$ & $3.09\times10^{59}$ & $3.15\times10^{44}$ \\
C2 & -- & --     & 0.0 & 0.25 &$4.1\times10^{9}$&$4.8\times10^{3}$& 31.0 & $8.46\times10^{58}$ & $2.69\times10^{59}$ & 0 & $3.61\times10^{44}$ \\
C3 & -- & --     & 0.0 & 0.25 &$9.5\times10^{8}$&$6.4\times10^{3}$& 31.0 & $2.64\times10^{59}$ & $6.23\times10^{58}$ & 0 & $3.33\times10^{44}$ \\
\hline
W0 & ~$7.9\times10^2$ & -- & -- & --& -- & -- & --& -- & -- & -- & -- \\
W1 & ~$7.9\times10^2$ & 15.0 & 3.0 & 0.10 &$4.6\times10^{7}$&0& 31.0 &  0 & $1.02\times10^{56}$ & $3.09\times10^{59}$ & $3.15\times10^{44}$ \\
W2 & ~$7.9\times10^2$ & --  & 0.0 & 0.25 &$4.1\times10^{9}$&$4.8\times10^{3}$&31.0 & $8.46\times10^{58}$ & $2.69\times10^{59}$ & 0 & $3.61\times10^{44}$ \\
W3 & ~$7.9\times10^2$ & --  & 0.0 & 0.25 &$9.5\times10^{8}$&$6.4\times10^{3}$& 31.0 & $2.64\times10^{59}$ & $6.23\times10^{58}$ & 0 & $3.33\times10^{44}$ \\
\hline
S0 & $^*$$7.9\times10^2$ & -- & -- & --& -- & -- & -- & -- & -- & -- & -- \\
S1 & $^*$$7.9\times10^2$ & 15.0 & 3.0 & 0.10 &$4.6\times10^{7}$&0& 31.0 &  0 & $1.02\times10^{56}$ & $3.09\times10^{59}$ & $3.15\times10^{44}$ \\
S2 & $^*$$7.9\times10^2$ & --  & 0.0 & 0.25 &$4.1\times10^{9}$&$4.8\times10^{3}$& 31.0 & $8.46\times10^{58}$ & $2.69\times10^{59}$ & 0 & $3.61\times10^{44}$ \\
S3 & $^*$$7.9\times10^2$ & --  & 0.0 & 0.25 &$9.5\times10^{8}$&$6.4\times10^{3}$& 31.0 & $2.64\times10^{59}$ & $6.23\times10^{58}$ & 0 & $3.33\times10^{44}$ \\
\hline
\end{tabular}
\end{center}
The initial capital letter in the model names represents the type of intra-cluster weather, ``C'' for the ``Calm'' cases (without wind or shock), ``W'' for the ``Wind'' cases, and ``S'' for the ``Shock'' cases, respectively; While the number in the model names represents the jet models, ``0'' for the cases without jet (control runs to study core sloshing), ``1'' for the cases with magnetic tower jet, ``2'' for the cases with subsonic hydrodynamical jet, and ``3'' for the cases with supersonic hydrodynamical jet, respectively. For the purpose of comparison, the AGN lifetime ($\Delta t_{\rm inj}$) is fixed to be 31 Myr, and the total jet power of each jet model is similar with each other, i.e., $\sim 3.3\times10^{44}$ erg/s.

$^*$ the wind speed in the shock cases is actually the initial velocity of the ICM behind the initial shock front, and the location of the initial shock front is set to be $x_{\rm sf}=-15$ (i.e., 45 kpc on the upstream from the galactic center, see \S\ref{sec:intra-cluster-weather} for more details).
\end{table*}

\subsection{Galactic Environment} \label{sec:icm-background}
In our simulations, we use stratified atmospheres to mimic the realistic environments of the head-tail radio sources. Taking 3C 465 for an example, there is a large X-ray halo with a size of several hundred kilo-parsec surrounding it, which is recognised as the core of its host cluster Abell 2634 \citep{hardcastle_chandra_2005}. The observations also show that there is a cold compact core in its host galaxy NGC 7720. The cold compact core is probably a consequence of weak cooling flow which could be supplying accretion fuel to the central supermassive black hole, so it may also be the reason why the galaxy is active \citep{schindler_x-ray_1997}. Motivated by the observations above, we initialise the jet's host galaxy and its cluster environment with a double-core structure --- a large smooth core of size $\sim 100$ kpc to mimic the intra-cluster environment, and a small compact core ($\sim 10$ kpc) to mimic the host galaxy. 
For each core, we use an isothermal King's beta model \citep{king_structure_1962},
\begin{equation}
\label{eq:kings-beta-model}
p = \rho = \rho_c \left[ 1+{(R/R_c)}^2\right]^{-\beta}
\end{equation} 
where $R=\sqrt{x^2+y^2+z^2}$ is the spherical radius, $R_c$ and $\rho_c$ are the core radius and typical core density, respectively. The parameter $\beta$ controls the gradient of the ambient medium. For the cluster core, the parameters are chosen as $R_c=4$, $\rho_c=1$ and $\beta=0.75$, while for the cold compact galactic core, $R_c=1$, $\rho_c=10$ and $\beta=1.0$.  

Furthermore, we assume the ambient gas is initially in hydrostatic equilibrium ---  the gravitational potential $\psi$ in Equation \ref{eq:momentum}-\ref{eq:energy} is initialised to satisfy the hydrostatic equilibrium condition given the density profile above. Then the gravity profile is fixed during the simulations \citep{nakamura_structure_2006}.

\subsection{Intra-Cluster Weather} \label{sec:intra-cluster-weather}

It is well accepted that the intra-cluster weather could have significant effects on the jet morphology as in the observed head-tail radio sources. Without loss of generality, two kinds of intra-cluster weather are studied in this paper: 1) wind (relative motion between the jet's host galaxy and its surrounding ICM) and 2) shock  (to mimic the cases of galaxy/cluster merger). 

To implement ICM winds in our simulations, we set inflow boundary conditions on the -x boundary, i.e., fixing the values of density and temperature in the boundary zone according to the ICM wind properties (magnetic field on the boundary is simply set to be zero), while giving a positive value to the x-component of the velocity $v_{\rm x}$. On the other boundaries, we use the outflow boundary condition to allow the flows freely in and out.

In the case of shock passage, we introduce the shock into our simulations by designing an initial condition, in which an initial positive $v_{\rm x}$ is given to the upstream ICM where $x<x_{\rm sf}$ {(rather than only at the boundary)}. $x_{\rm sf}$ is a free parameter, which we call the ``initial  shock front'' (see Table \ref{tab:runs} for more details). The density and temperature are unchanged in the initial setup. {Unlike the wind case, the values of $v_{\rm x}$, density and temperature will change as the flow evolves}. Finally, all of the boundaries are set with outflow boundary conditions. This will naturally induce sound/shock wave passing across the jet.

The most uncertainty in our model lies on the ICM wind properties. Generally speaking, the density of the winds should smoothly connect to the ICM density on the outskirt of the galaxy. In our simulations, the wind density and temperature are chosen as their initial values at the boundary. Observationally, the wind speed is very difficult to be determined.  However, we could still make some rough estimation of the wind speed from the observed morphology of head-tail galaxies.  Taking 3C 465 for an example, the length of the tails is $\sim 100$ kpc. If we assume a typical lifetime of 100 Myr for the radio jet, then the wind speed required to bend the jet, before it fades off, should be $\sim 980$ km/s. It implies that the wind speed should be transonic at least.

Finally, we build cartesian coordinates co-moving with the jet central engine. When without notation, the computational domain is usually taken to be $|x|\le56$, $|y|\le56$, and $|z|\le56$, which corresponds to a box of $(560\ {\rm kpc})^3$ in actual length scales. The numerical resolution is chosen to be $512^3$, and we use a non-uniform mesh to resolve the central region better, where the smallest grid size is set to be 0.0625. We've also done the resolution study with doubled mesh size in each direction, we find that our conclusions are unchanged.

\section{Results} \label{sec:results}
In this section, we perform detailed analysis on the jet bending process under violent intra-cluster weather. 
Before we get into the detailed physics, it is helpful to make some generic considerations about the jet morphology, wind properties and the intra-cluster ICM environment. 
First of all, to get a general picture of the morphology of head-tail sources, we take 3C 465 (the WAT prototype, Figure \ref{fig:cartoon-head-tail-sources}, left panel) as an example: there are two arms in its radio image, each of the arms has a ``slim'' straight part of a length $\sim 30-50$ kpc attached to the galactic center, a fat/straight part of $\sim 100$ kpc, a twist ribbon of $\sim 150$ kpc and a broad big fluffy tail of $\sim 200$ kpc \citep[see also][]{eilek_magnetic_2002}. 
As we argue in \S\ref{sec:intra-cluster-weather}, it would be a good starting point by assuming a transonic wind, of which the speed $\sim 1000$ km/s (for a typical ICM temperature of 4 kev). Given the tail length of $\sim 200$ kpc, we could immediately get the timescale needed to bend the jet, $\sim 200$ Myr. For a generic consideration of a typical AGN lifetime of tens of Myr, it is likely that the jet bending process occurs mainly {when the jet engine already turns off.}

The plan of this section is as follows: to isolate the physical effect between the jet, ICM, and wind, we study the wind-ICM interaction in \S\ref{sec:core-sloshing} (we find core sloshing is inevitable, which is very important to the final jet morphology) and the jet-ICM interaction in \S\ref{sec:static-jet} (we give some references of the jet behaviour without intra-cluster weather), respectively, before we focus on the detailed analysis of the jet bending process in \S\ref{sec:jet-bending}.  The model parameters of all the simulation runs in this paper are listed in Table \ref{tab:runs}.

\subsection{Core Sloshing} \label{sec:core-sloshing}

In this subsection, we study the wind-ICM interaction via a control numerical model (Run W0) as described in \S\ref{sec:model-setup} but without a jet, the model parameters are listed in Table \ref{tab:runs}. As described in \S\ref{sec:icm-background}, we assume that the ICM environment is in hydrostatic equilibrium when we set up the initial ICM profile. 
When the wind comes, the ICM could gain some speed by absorbing kinetic energy from the incident wind material, and deviates from the initial equilibrium status. If the velocity of the disturbed ICM is larger than the local escape velocity, it will be directly stripped off from the cluster. 
As a result of the competition between the wind ram pressure and cluster gravity, core sloshing is inevitable after the cluster core absorbs enough kinetic energy from the wind. 

\begin{figure*}[t]
\label{fig:core-sloshing-phases}
\centering
\includegraphics[width=0.9\textwidth]{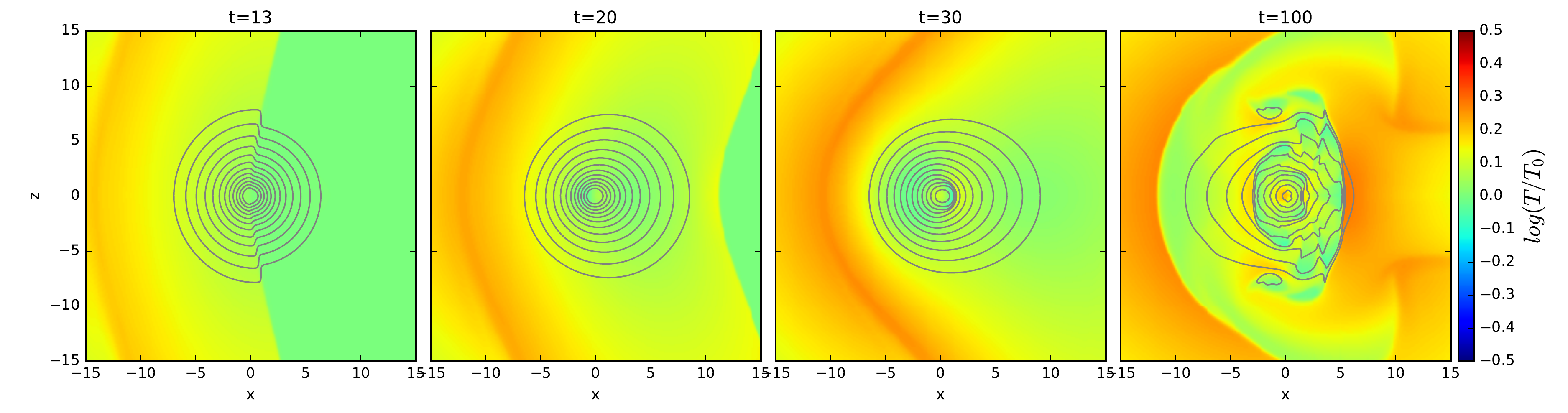}
\includegraphics[width=0.9\textwidth]{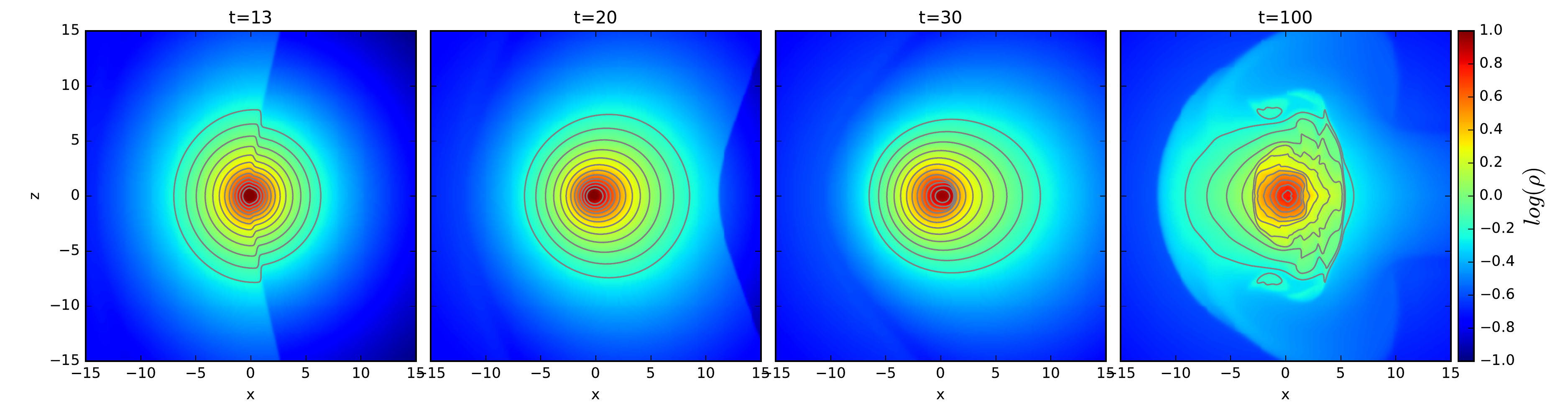}
\includegraphics[width=0.9\textwidth]{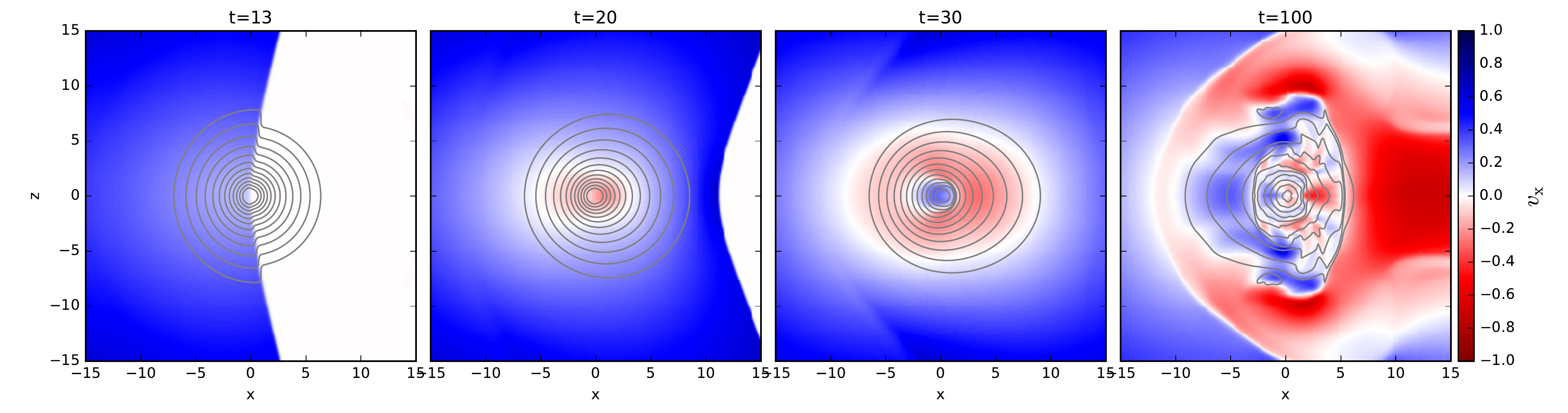}
\caption{Core sloshing driven by wind (Run W0).  Each panel, from top to bottom, shows snapshots (sliced from 3D data at $y=0$) of temperature, density, and x-component of the ICM velocity in the core region at time $t=13,20,30,100$, respectively. While the grey lines are contours of normalized Bremsstrahlung emissivity ($j_{\rm brem} = \rho^2\sqrt{T}$).}
\end{figure*}

 
In Figure \ref{fig:core-sloshing-phases}, we present snapshots of temperature (upper panel), density (middle panel), and the x-component of the fluid velocity $v_{\rm x}$ (lower panel) in the core region at time $t=13,20,30,100$, respectively. The grey lines are contours of Bremsstrahlung emissivity. We can see clearly two fronts propagating in x-direction, i.e., (1) the shock front driven by the ICM wind (we call it the ``wind shock'' hereafter), as shown in the columns of the figure, which passes through the core and crosses the points (x,y,z) $\simeq$ (0,0,0), (11,0,0) at $t=13,20$, respectively; and (2) the contact discontinuity surface between the wind material and ICM, which crosses the points (x,y,z) $\simeq$ (-14,0,0), (-12,0,0), (-10,0,0), (-12,0,0) at $t=13,20,30,100$, respectively. Unlike the wind shock, the wind material could not penetrate through the core region as argued above, instead, the kinetic energy carried by wind is absorbed by the ICM core as the material is bound to the system despite the strong disruption.

\begin{figure}[t]
\label{fig:vx-in-zt-plane}
\centering
\includegraphics[width=0.5\textwidth]{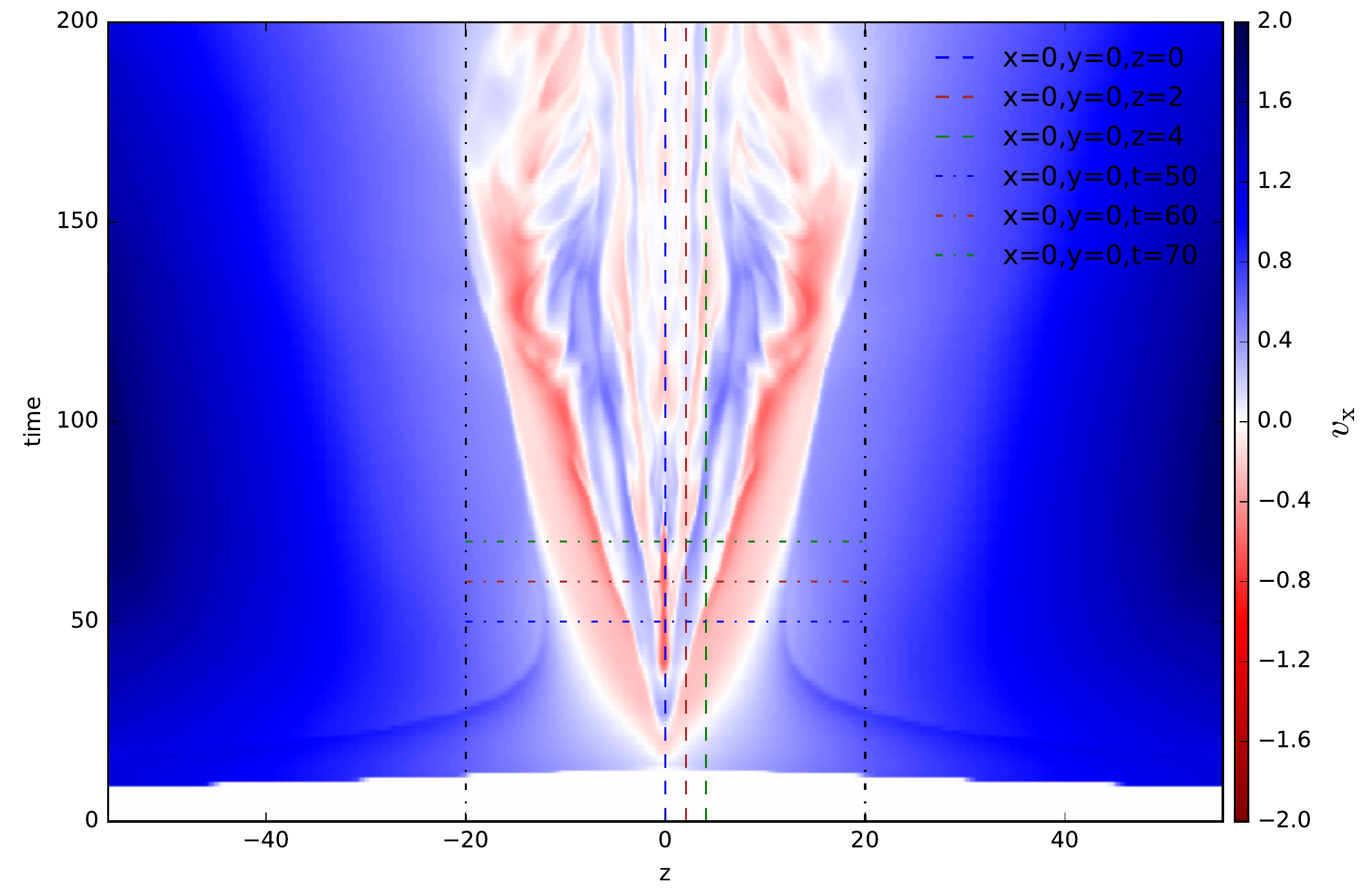}
\caption{Pseudocolor map for $v_x$ in the z-time plane (Run W0). It shows the time variation of the values of $v_{\rm x}$ along the z-axis ($x=0$, $y=0$). The vertical dash-dotted lines indicate the boundaries of the bound/unbound ICM. 
Beyond the boundaries, the ICM is stripped off the cluster (non-negative $v_{\rm x}$ along the time axis), 
while between the boundaries is the core sloshing region as a result of the competition between the impact of wind and cluster gravity. To give more quantified details, we make some slices from this figure (the horizon dash-dotted lines and the vertical dashed lines), which are shown in Figure.\ref{fig:vx-variation}.}
\end{figure}

In Figure \ref{fig:vx-in-zt-plane}, we show the time variation of the values of $v_{\rm x}$ along the z-axis ($x=0$, $y=0$). The vertical dash-dotted lines (at $z\sim\pm20$) indicate the boundaries of the bound/unbound ICM. As we argue above, beyond the boundaries the ICM is directly stripped off the cluster (non-negative $v_{\rm x}$ along the time axis), while between the boundaries is the core sloshing region, where the cluster gravity competes with the wind ram pressure.

\begin{figure}[t]
\label{fig:vx-variation}
\centering
\includegraphics[width=0.45\textwidth]{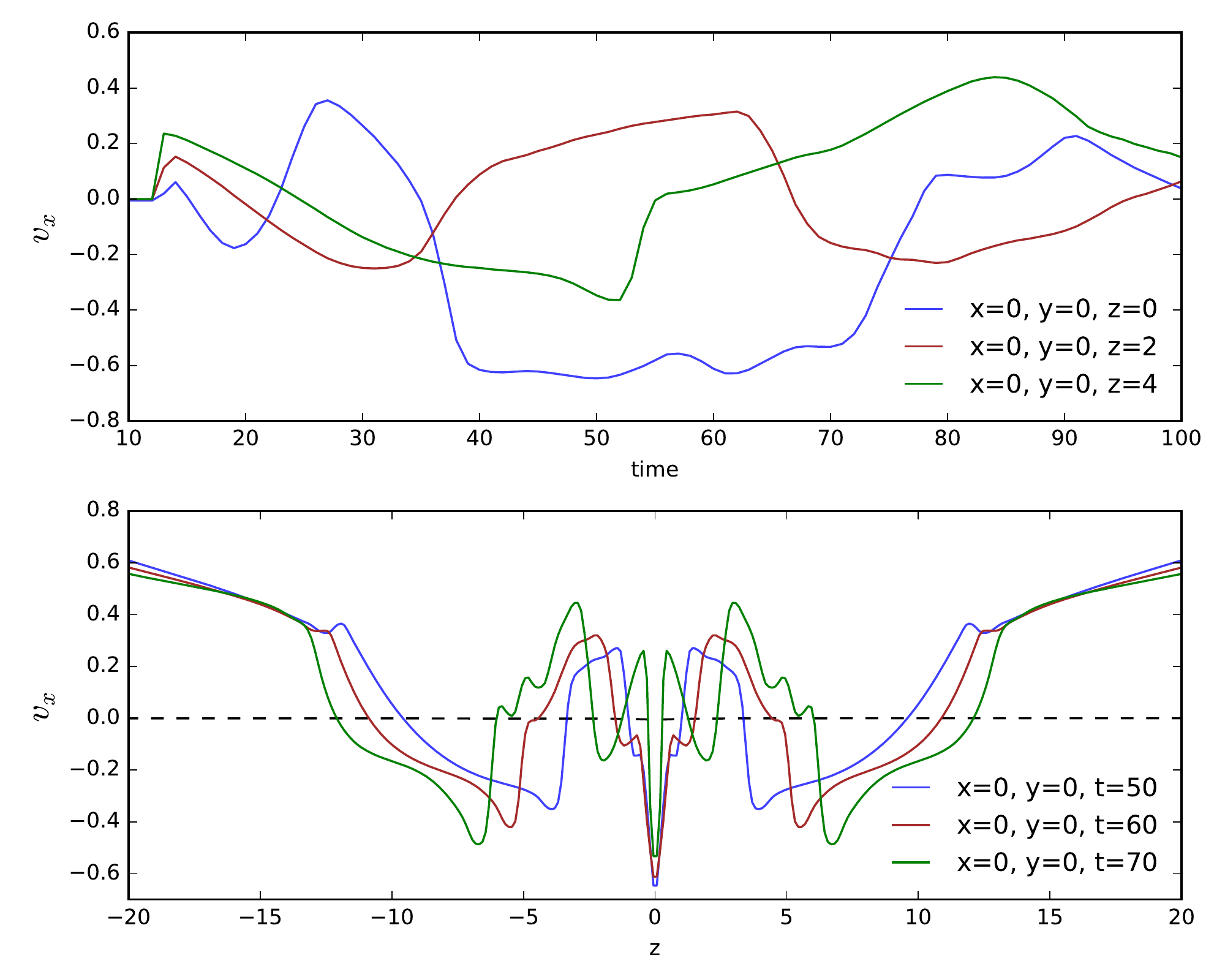}
\caption{The asynchronous sloshing patterns and the strong shear motion along the z-axis (Run W0). 
The data are sliced from Figure \ref{fig:vx-in-zt-plane} --- the lines in the upper panel correspond to the vertical dashed lines in Figure \ref{fig:vx-in-zt-plane}, i.e. the time variation of $v_{\rm x}$ at some specific locations along the z-axis,which shows the asynchronous sloshing patterns for different radius; while the lines in the lower panel correspond to the horizon dash-dotted lines in Figure \ref{fig:vx-in-zt-plane}, i.e. the profiles of $v_{\rm x}$ along the z-axis at some specific times, which shows the strong shear motion (along the z-axis) induced by the asynchronous core sloshing.}
\end{figure}

\begin{figure*}[t]
\label{fig:richardson-number}
\centering
\includegraphics[width=0.9\textwidth]{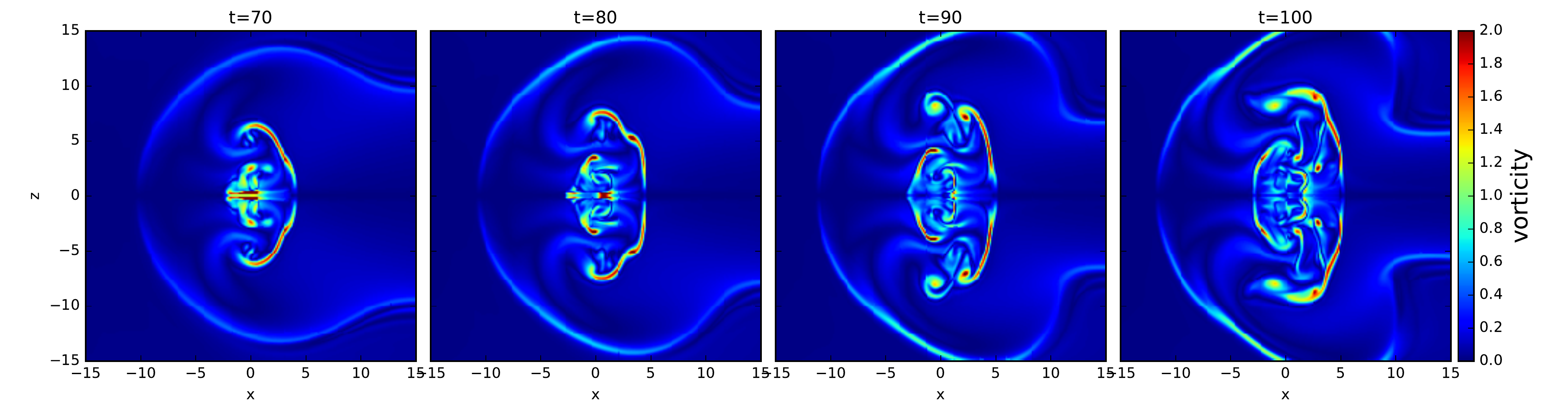}
\includegraphics[width=0.9\textwidth]{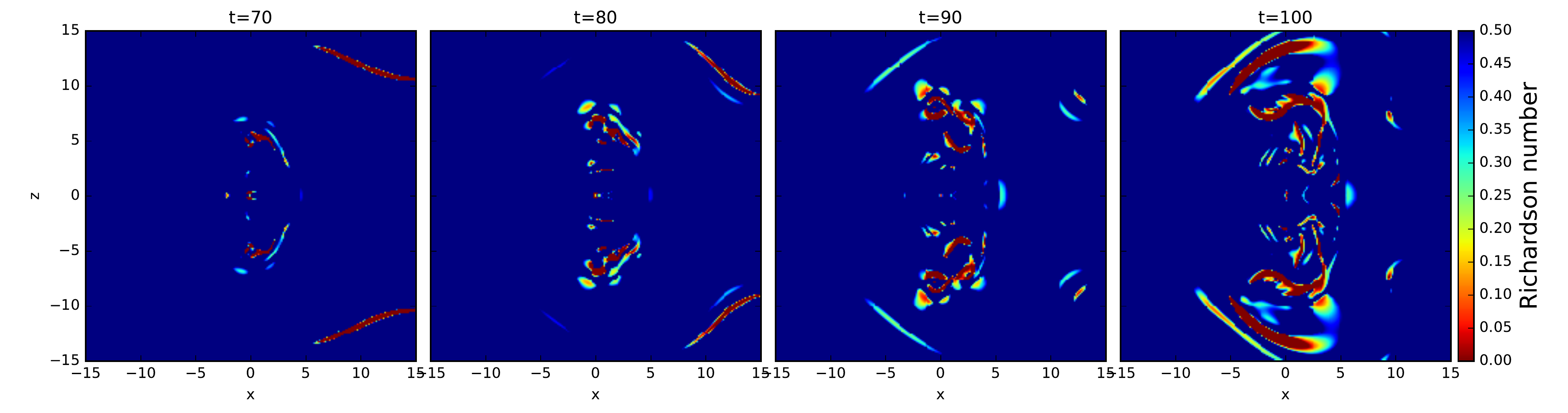}
\caption{Kelvin-Helmholtz instability found in the late stage of core sloshing (Run W0). the upper panel shows the snapshots of vorticity $w=|curl(v)|$ at four selected time points (from left to right, $t=70,80,90,100$, respectively), the lower panel shows the results of the Richardson number $\mathrm{Ri}$ calculated from Equation \ref{eq:richardson-number} . It is shown that the turbulent region matches well with the region of $\mathrm{Ri}<1/4$, when compared to Figure \ref{fig:core-sloshing-phases}. }
\end{figure*}

We can see clearly from Figure \ref{fig:core-sloshing-phases}-\ref{fig:vx-in-zt-plane} that the core sloshing firstly undergoes a linear stage (laminar-like flows, when $t<\sim70$), and finally excites turbulence around the cluster core radius (when $t>\sim100$). Roughly speaking, the wind-ICM interaction process could be divided into three phases in the order of time: 
\begin{enumerate}
\item
Impacting phase (as shown in the leftmost column of Figure \ref{fig:core-sloshing-phases}). 
In this phase, the wind shock begins to pass through the core region. The ICM gains momentum from the shock, and moves forward in the wind direction. The duration of this phase is relatively short, the ICM motion in the innermost region, where the dynamical timescale is the shortest, usually quickly switches into the sloshing phase shortly after the shock front passes it. While for larger radii, there is a relative time delay before it becomes sloshing (see also Figure \ref{fig:vx-in-zt-plane}, at $t>13$).

\item
Sloshing phase. 
The competition between the wind ram pressure and the cluster gravity results in core sloshing in the direction of the wind (i.e., the x-direction). Consequently, $v_{\rm x}$ of the sloshing ICM  keeps changing its sign as shown in Figure \ref{fig:vx-in-zt-plane} (e.g.,  along the vertical dashed lines). In Figure \ref{fig:vx-variation}, we present more details about the variation of $v_x$ at specific locations (the upper panel) and at specific times (the lower panel), which are extracted from Figure \ref{fig:vx-in-zt-plane} along the vertical dashed lines, and the horizontal dash-dotted lines, respectively. It is shown in the upper panel of Figure \ref{fig:vx-variation} that $v_x$ at $z=0,2,4$ changes its sign for the first time within a time interval of $\Delta t\simeq4,8,13$, respectively, i.e., the sloshing pattern for different radius is {\it asynchronous}. Since the dynamical timescale is the shortest in the innermost region, core sloshing firstly appears there (see the second column of Figure \ref{fig:core-sloshing-phases}). Eventually, the whole core region becomes sloshing, and the sloshing amplitude increases with time as the core absorbs more and more kinetic energy from the wind. In this phase, we could clearly see the elongated core structure (see the third column of Figure \ref{fig:core-sloshing-phases}, the grey contours of Bremsstrahlung emissivity), which is very similar to the observed X-ray image of 3C 465 \citep[see, e.g.,][]{eilek_magnetic_2002}.  

\item
Turbulence phase. 
As shown in the lower panel of Figure \ref{fig:vx-variation}, significant shear motion in $v_{\rm x}$ could be induced along the z-axis because of the asynchronous core sloshing. In Figure \ref{fig:richardson-number}, we present snapshots of vorticity (the upper panel) and the Richardson number $\mathrm{Ri}$ (the low panel) at four selected time points (from left to right, $t=70,80,90,100$, respectively). It is shown that, when the amplitude of core sloshing becomes high enough (after the core absorbs enough kinetic energy from the wind), the threshold ($\mathrm{Ri}<1/4$) could be reached for Kelvin-Helmholtz instability to grow up, and turbulence is finally excited around the core region (see the rightmost column of Figure \ref{fig:core-sloshing-phases}).  
\end{enumerate}

The Richardson number $\mathrm{Ri}$ shown in Figure \ref{fig:richardson-number} is defined as
\begin{equation}
\label{eq:richardson-number}
\mathrm{Ri} 
                       = -\frac{\nabla\psi}{\rho} \frac{\nabla \rho}{(\nabla v)^2}.
\end{equation}
\noindent
Theoretically, the velocity shear could overcome the tendency of a stratified fluid to remain stratified when $\mathrm{Ri}<1/4$, and some mixing (e.g., Kelvin-Helmholtz instability in our situation) will generally occur. We can see clearly, from the lower panel of Figure \ref{fig:richardson-number},  that the turbulent region is well coupled with the region of $\mathrm{Ri}<1/4$ (red-colored), i.e., it is the Kelvin-Helmholtz instability (driven by core sloshing) excites turbulence in the core region. 

The duration of the impacting phase, sloshing phase and turbulence phase are approximately 4 (25 Myr), 54 (335 Myr), and $>130$ (806 Myr), respectively, in our fiducial model (Run W0) as shown in Figure \ref{fig:core-sloshing-phases}-\ref{fig:richardson-number}. As argued at the beginning of this section, the time needed to bend the jet (like the one of 3C 465) is in the order of 100 Myr, which is much longer than the duration of the impacting phase. So, {\it core sloshing is inevitable during the jet bending process}.

We expect that the effects of core sloshing could be twofold: on the one hand, it is a challenge for jet to survive, since the ICM density is high in the core region, light jet could be easily twisted or even destructed by the shearing medium; On the other hand, {it could also help the jet to homogenise its energy into the ICM \citep{Heinz_answer_2006}.}

\subsection{Jet Propagation in Static Environment} \label{sec:static-jet}

\begin{figure*}[t]
\label{fig:static-jet}
\centering
\includegraphics[width=0.325\textwidth]{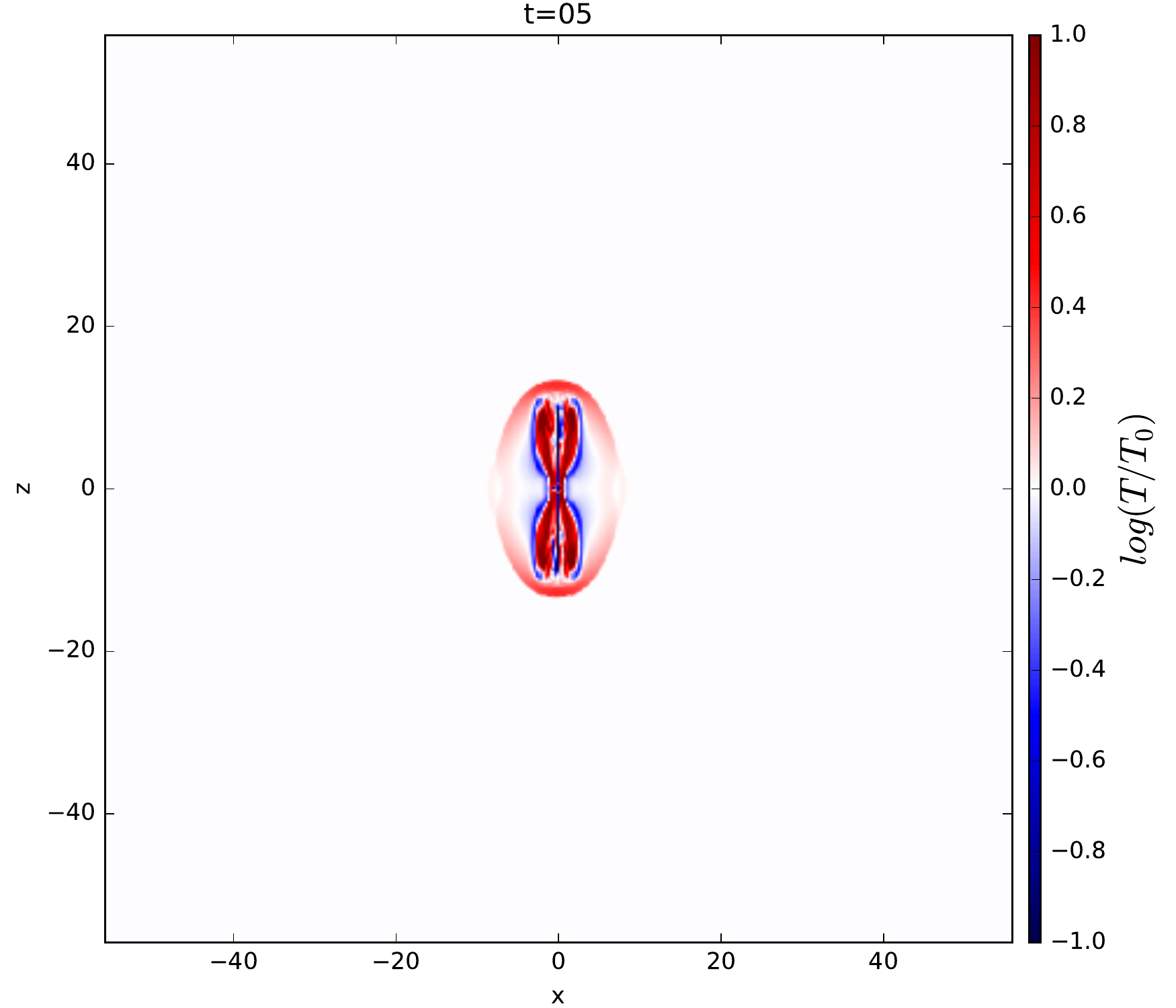}
\includegraphics[width=0.325\textwidth]{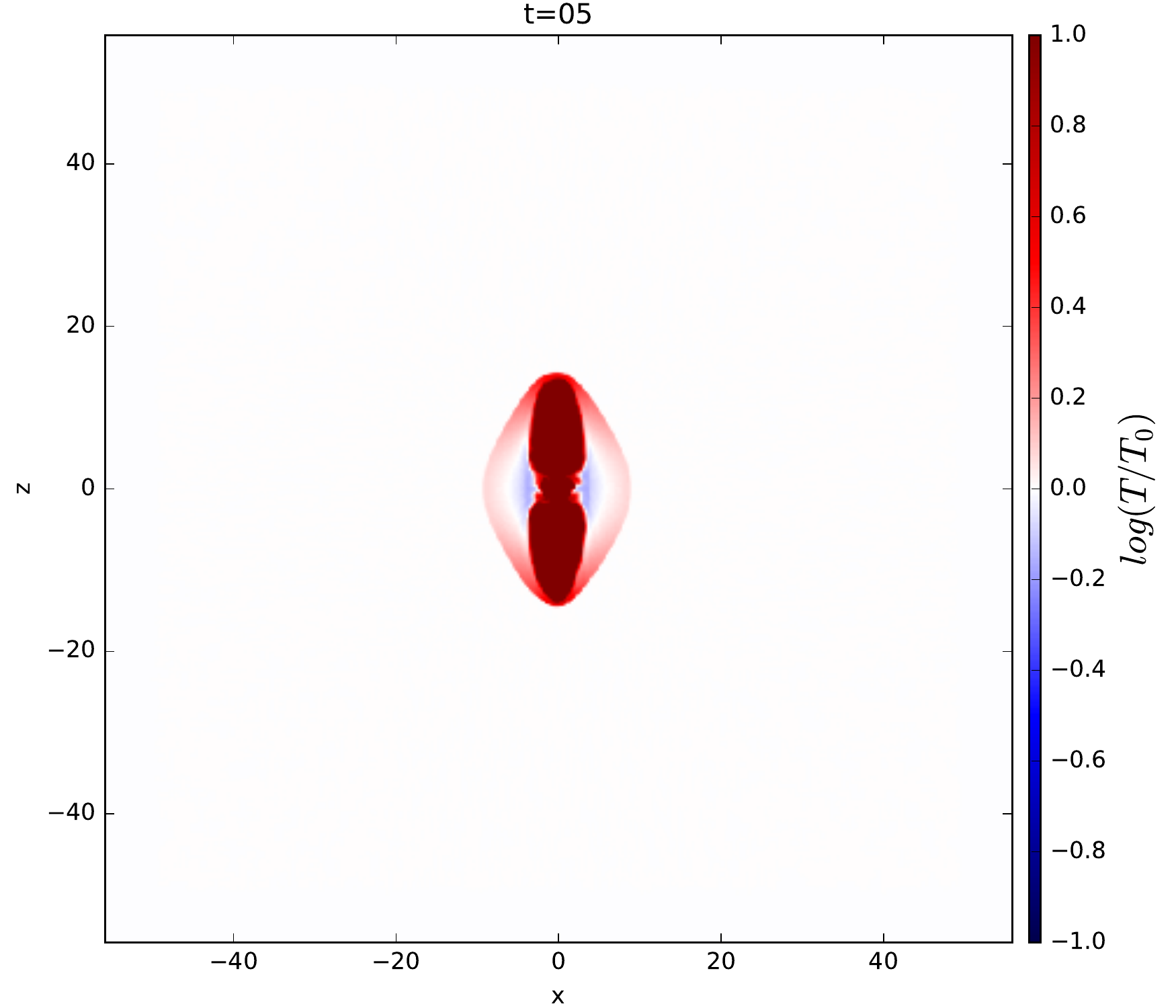}
\includegraphics[width=0.325\textwidth]{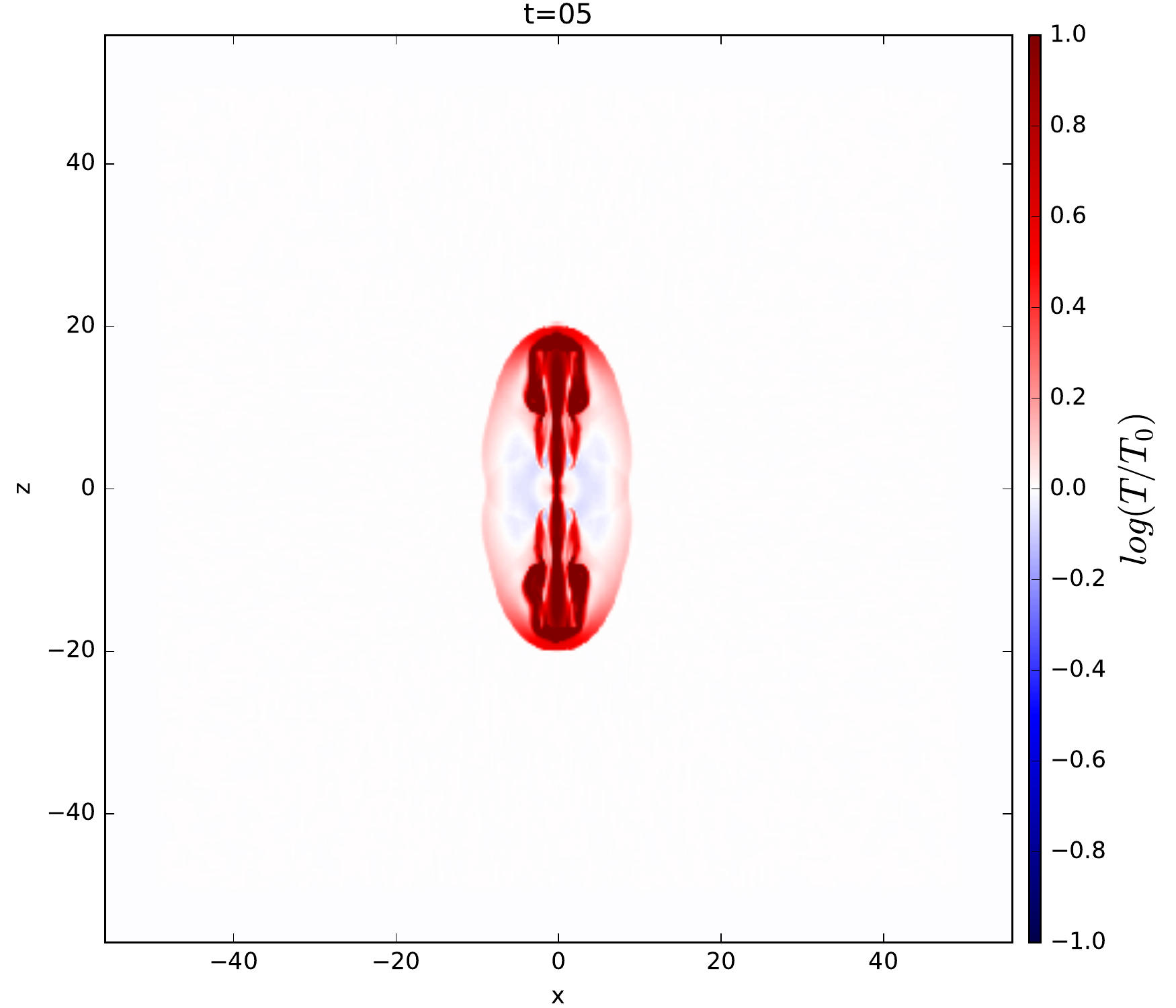}
\includegraphics[width=0.325\textwidth]{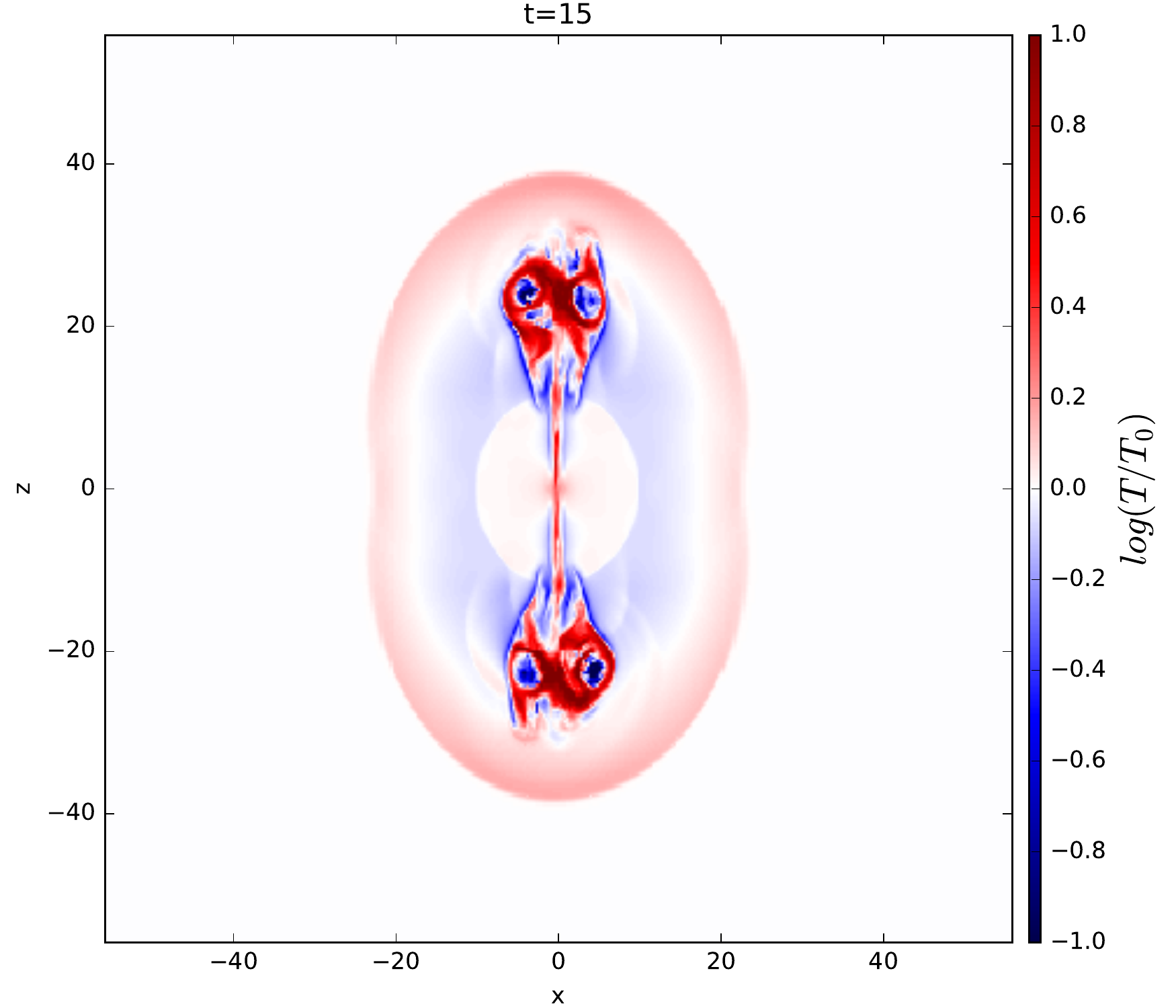}
\includegraphics[width=0.325\textwidth]{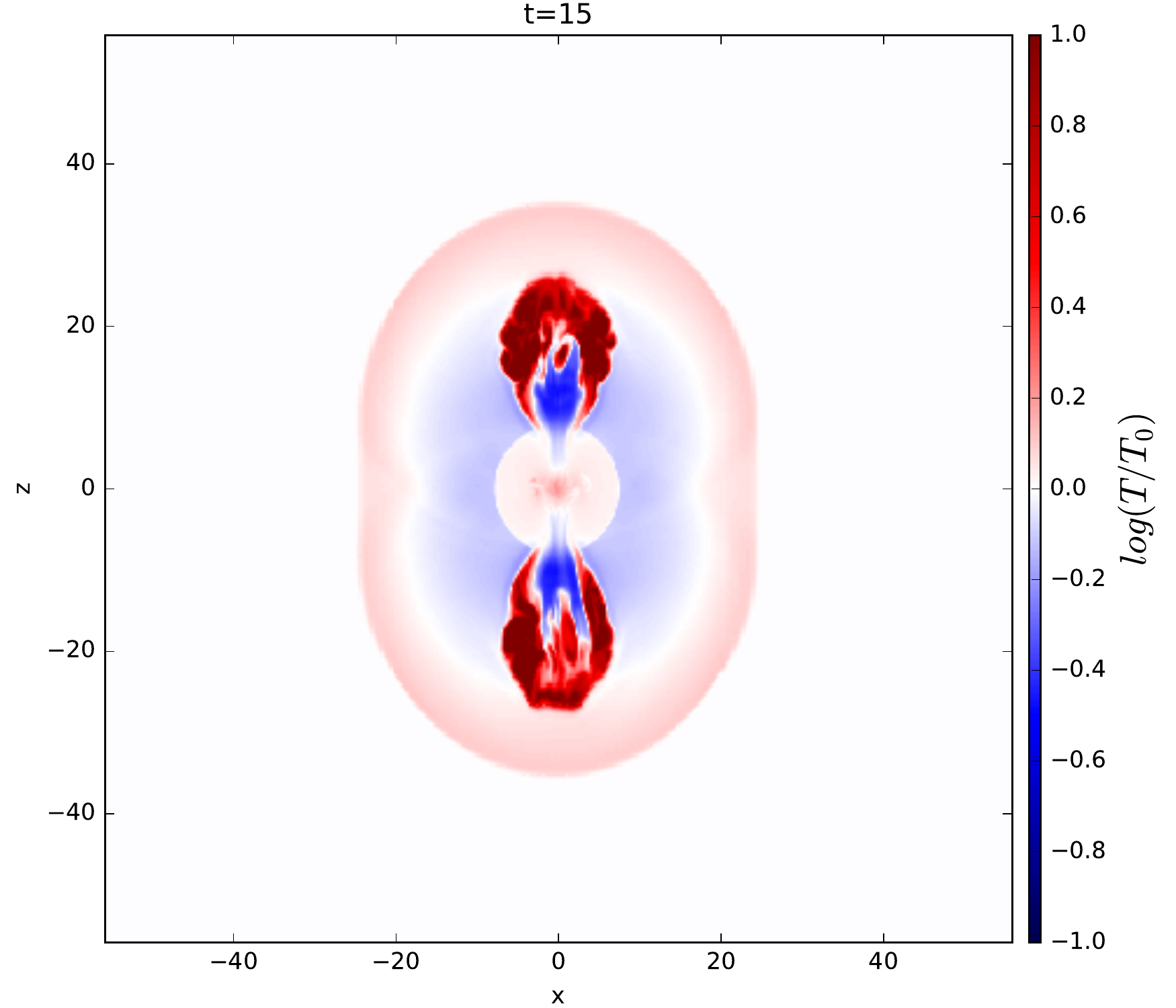}
\includegraphics[width=0.325\textwidth]{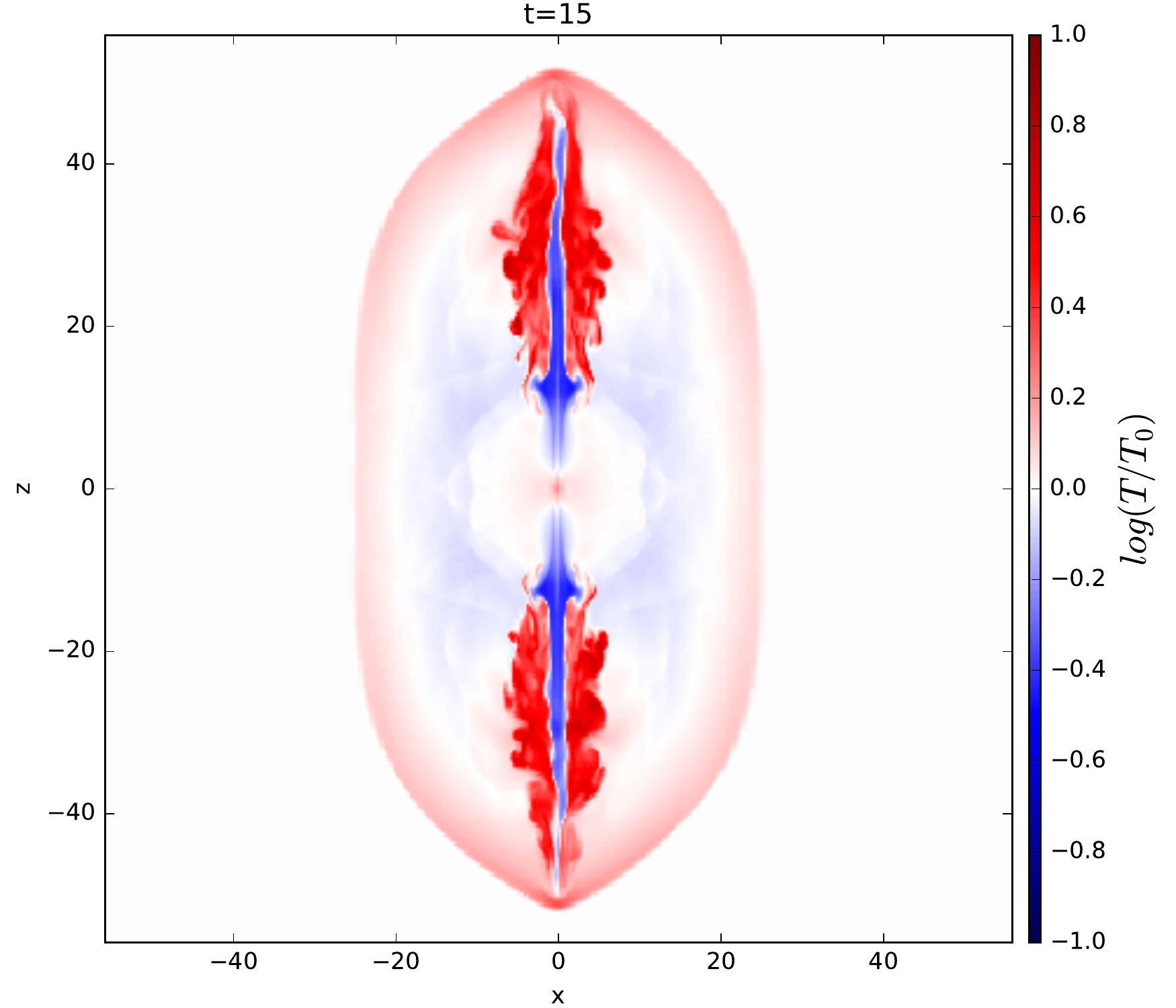}
\includegraphics[width=0.325\textwidth]{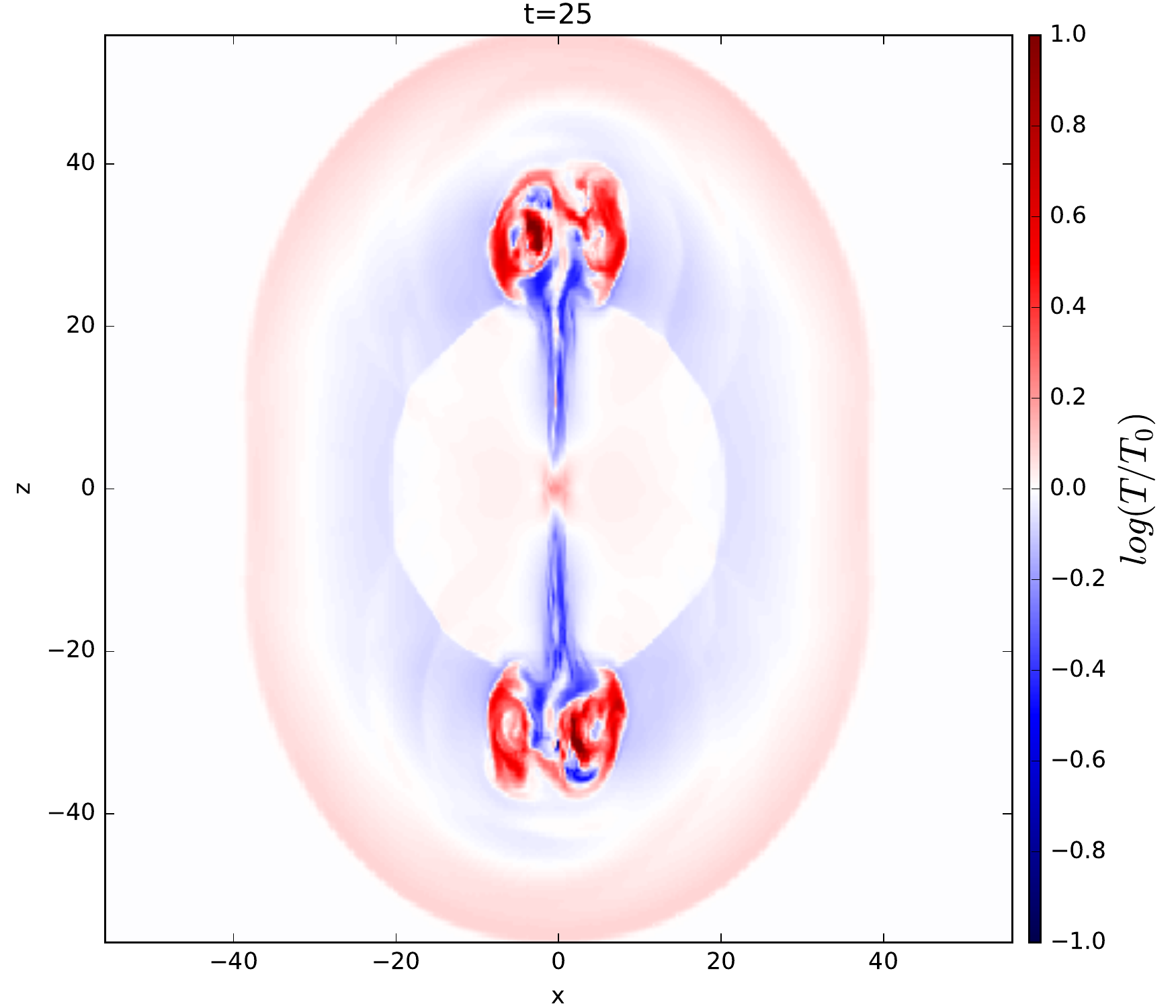}
\includegraphics[width=0.325\textwidth]{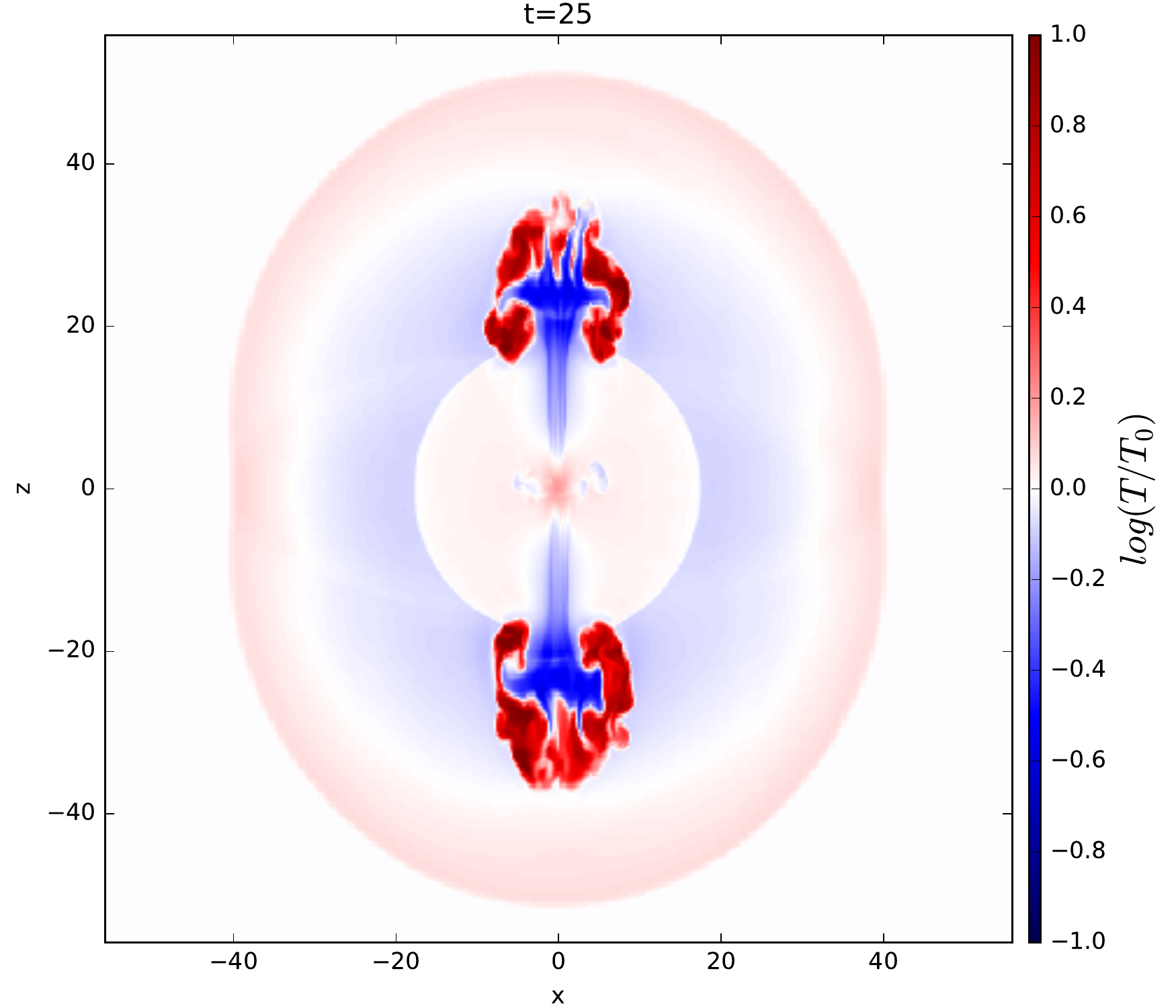}
\includegraphics[width=0.325\textwidth]{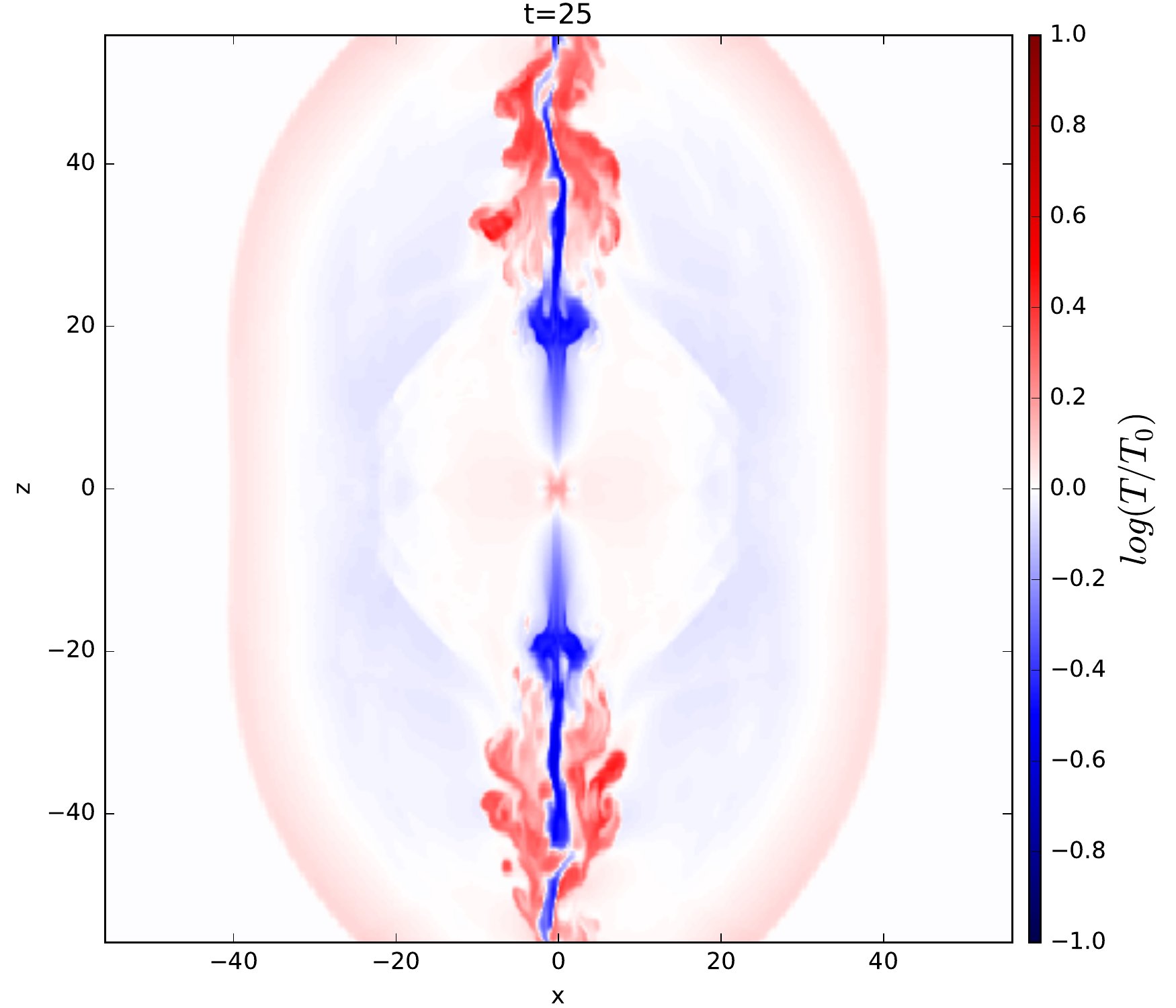}
\caption{Jet propagation in the static cluster environment. From left to right panels, the figures show the pseudocolor map of temperature for the cases of magnetic tower jet (Run C1), subsonic hydrodynamical jet (Run C2), supersonic hydrodynamical jet (Run C3), respectively. The upper, middle and lower panels are the results in a time sequence of $t=$ 5, 15 and 25, respectively. }
\end{figure*}

To isolate the process of jet-lobe formation, and also for the completeness of this paper, in this subsection we perform another series of control simulations on the jet propagation in the static environment (i.e., no wind or shock present). 
The lifetime of jet engine is fixed to be 5 (31 Myr) in all of our jet models. For the purpose of comparison, the jet parameters are fine tuned so that the jet power in every model is similar to each other, i.e. $\sim 3.3\times10^{44}$ erg/s, which corresponds to a total jet energy $\sim 3.2\times10^{59}$ erg. As listed in Table \ref{tab:runs}, Run C1 is for the magnetic tower jet, in which the magnetic energy overwhelmingly dominates over the kinetic energy and internal energy at injection. The mach number $\mathcal{M}$ of the magnetic tower jet in Run C1 is $\sim 0.85$. While in Run C2 (the subsonic hydrodynamical jet, $\mathcal{M}\sim0.65$), the internal energy dominates over the kinetic energy (and the magnetic field is simply set to be zero). Finally, it is dominated by kinetic energy in Run C3 (the supersonic hydrodynamical jet, $\mathcal{M}\sim2.60$).
The results are presented in Figure \ref{fig:static-jet}. From left to right, the columns of figures show the pseudocolor map of temperature for Run C1, C2 and C3, respectively. The upper, middle and lower panels are the results in a time sequence of $t=$ 5, 15 and 25, respectively.


It is well known that hydrodynamical jet suffers the Kelvin-Helmholtz and Rayleigh-Taylor instabilities, which are induced by its significant shear motion relative to the ambient ICM and the deceleration effect during the jet-ICM interaction, respectively.  As shown in the right column of Figure \ref{fig:static-jet}, a lot of small eddies, as a result of the Kelvin-Helmholtz instability, are induced along the path of the supersonic hydrodynamical jet (Run C3). The speed is so high that the supersonic jet usually tends to open a channel through which most of the jet energy leaks out of its host galaxy \citep{vernaleo_agn_2006}, and there isn't enough time for the jet to form lobes via adiabatic expansion. In the case of subsonic hydrodynamical jet (Run C2, middle column of Figure \ref{fig:static-jet}), the jet internal temperature is very high, and large lobes are allowed to form during the jet propagation. However, the lobes are found to be unstable again because of the hydrodynamical instabilities, and could not keep its integrity during the buoyant raising process (see also, e.g., \citealp{reynolds_buoyant_2005} and references therein).

The propagation of magnetic tower jet (Run C1) is quite different from that of hydrodynamical jets. As it is known that magnetic fields could suppress hydrodynamical instabilities, the surface of the magnetic tower jet is quite smooth when compared to the hydrodynamical jets (as shown in the left column of Figure \ref{fig:static-jet}). Instead, such a magnetic tower jet suffers the internal kink instability which usually occurs at places where there is some sudden change in its surrounding environment, e.g. at the cluster core radius where ICM pressure drops rapidly as radius increases. It turns out that the kink instability could disturb the helix structure of the magnetic fields and brake the jet propagation, then large lobes (driven mainly by the internal current) are allowed to form. Also, some of the``nodes" induced by the kink instability could be further compressed by the succedent jet material, which could induce some filamentary sub-structures (current sheets) within the lobes (see Figure \ref{fig:tower-jet-lobe}). Because of the global nature (divergence free) of magnetic field and the plasma frozen-in effect, the internal magnetic fields could protect the jet material from being dispersed. Under such protection, the integrity of the jet morphology is guaranteed for magnetic tower jet (for more details on the dynamics and instabilities of magnetic tower jet, please see \citealp{li_modeling_2006,nakamura_stability_2007}).


\subsection{Jet Bending Under Violent Intra-Cluster Weather} \label{sec:jet-bending}

\subsubsection{Jet Bent by Wind} \label{sec:jet-wind-interaction}
In this subsection, we perform detailed analysis on the jet bending process under violent intra-cluster wind. 
The model setup for jet and ICM environment is exactly the same as those in \S\ref{sec:static-jet} except that we use an inflow boundary condition at the -x boundary ($x=-20$) to allow ICM wind flowing into the computational domain.  The wind is of uniform profile on the -x boundary, i.e., with constant density, temperature, velocity and zero magnetic intensity (see \S\ref{sec:intra-cluster-weather} and Table \ref{tab:runs} for more details).

\begin{figure*}[t]
\label{fig:tower-jet-bending}
\centering
\includegraphics[width=0.9\textwidth]{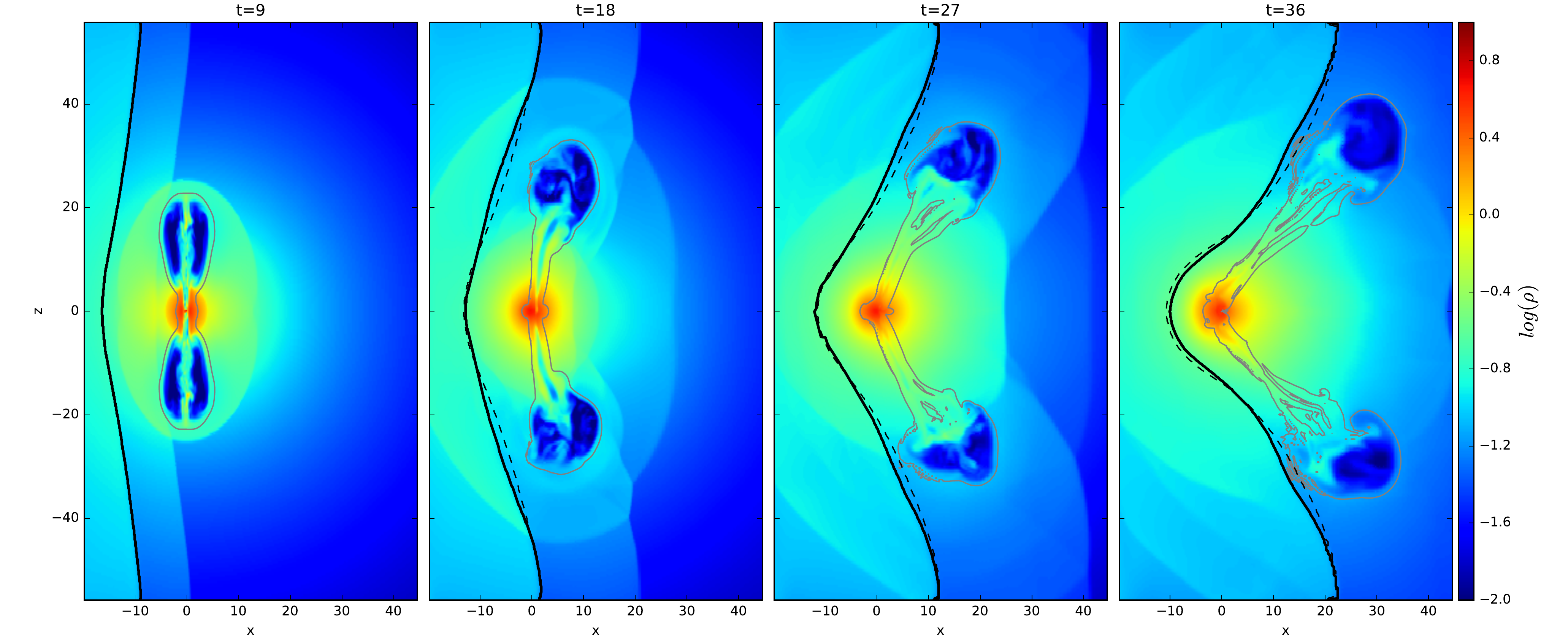}
\includegraphics[width=0.9\textwidth]{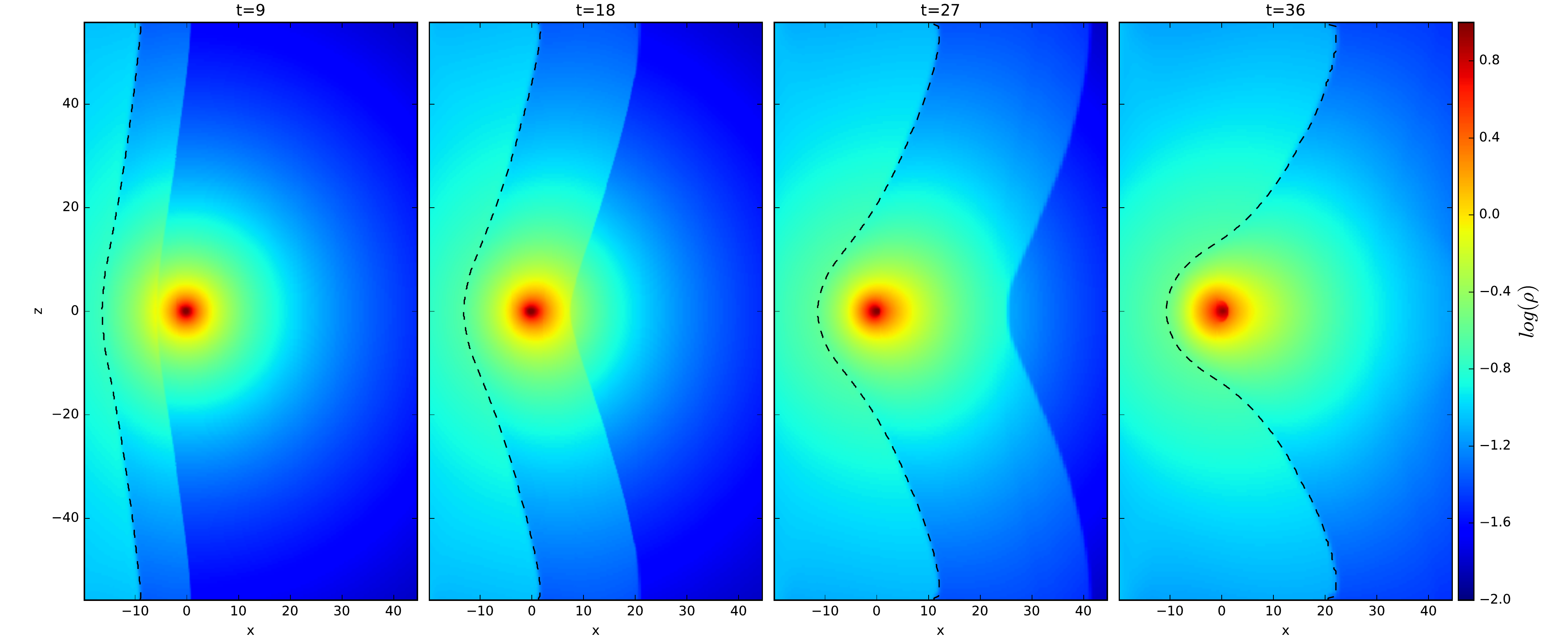}
\caption{Bending process of a magnetic tower jet (Run W1, the upper panel). The pseudocolor map in each figure shows the density profile (sliced from 3D data at $y=0$). The grey lines in the upper panel are the contours of magnetic intensity, which is used to indicate the jet morphology, while the black-thick-solid line is the contact discontinuity surface between the wind and ICM. The lower panel is for the case of Run C1 --- the dashed line is the contact discontinuity surface between the wind and ICM, while the dashed lines in the upper panel are just taken from the lower panel for the convenience of comparison. {\bf It is shown that the wind-ICM CD surfaces almost superimpose completely with each other for the cases with and without a jet. The opening angle of the wind-ICM CD surfaces are about $160^{\rm o}$, $145^{\rm o}$, $125^{\rm o}$, $105^{\rm o}$ at $t=$ 9, 18, 27, 36, respectively.}}
\end{figure*}
\begin{figure*}[t]
\label{fig:tower-jet-bending-suppl}
\centering
\includegraphics[width=0.75\textwidth]{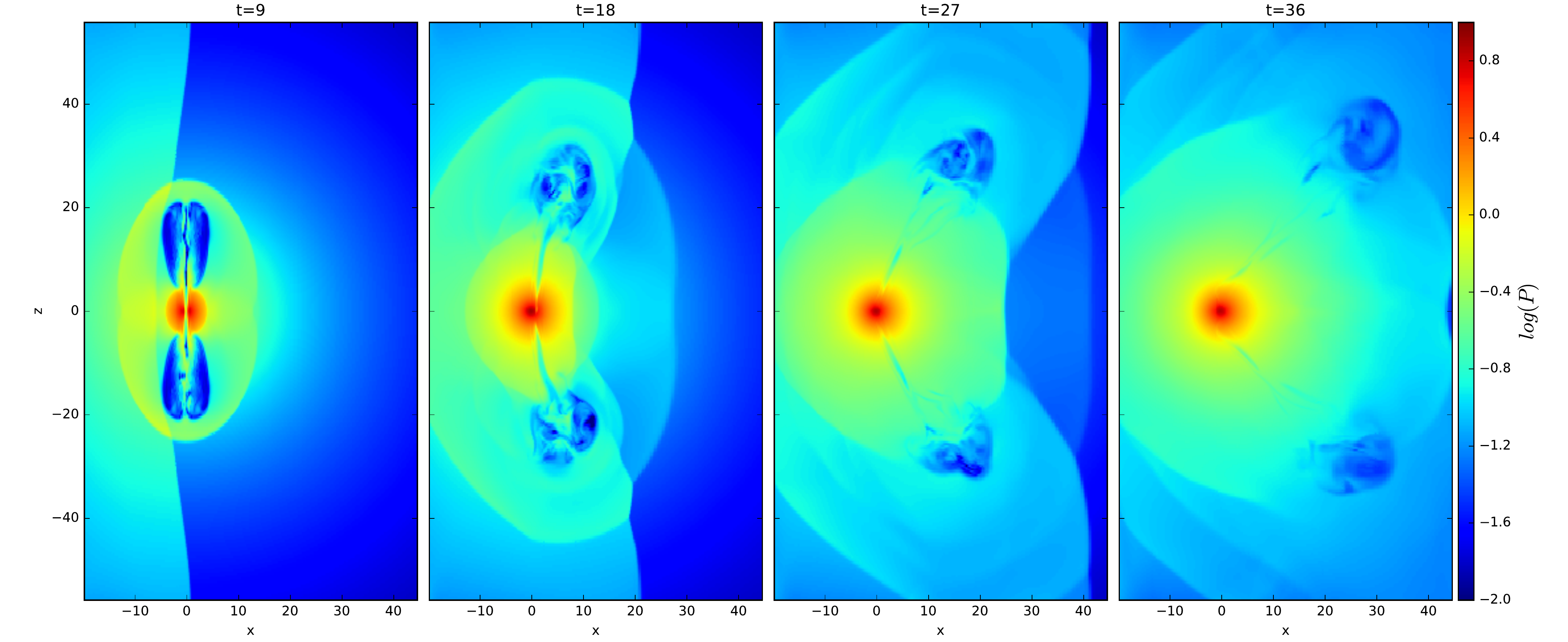}
\includegraphics[width=0.75\textwidth]{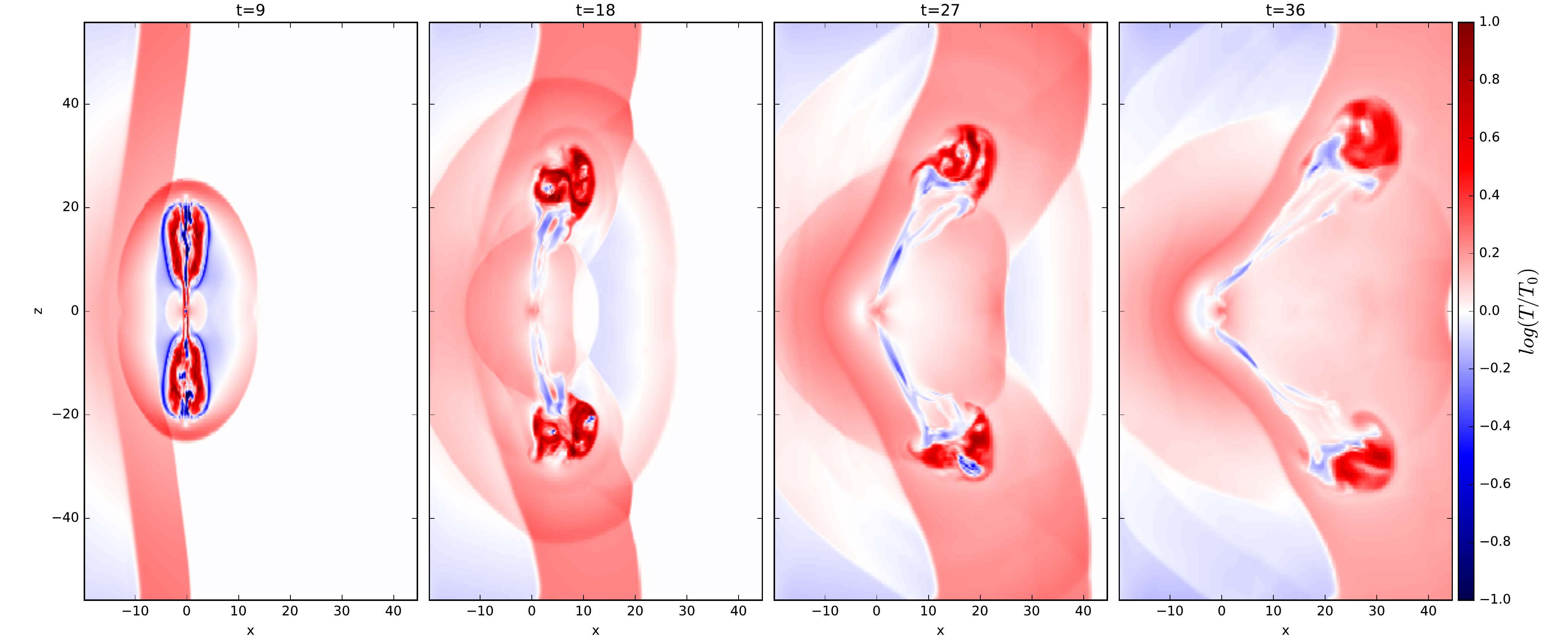}
\includegraphics[width=0.75\textwidth]{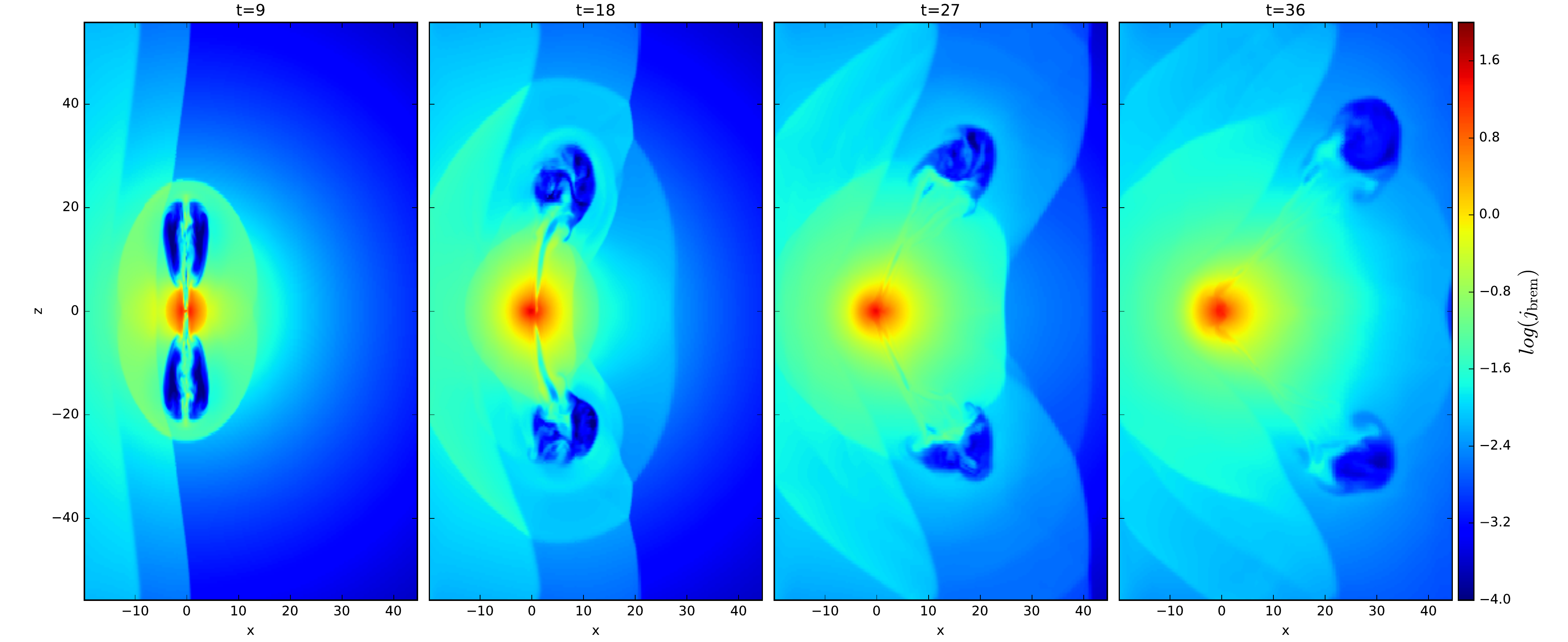}
\includegraphics[width=0.75\textwidth]{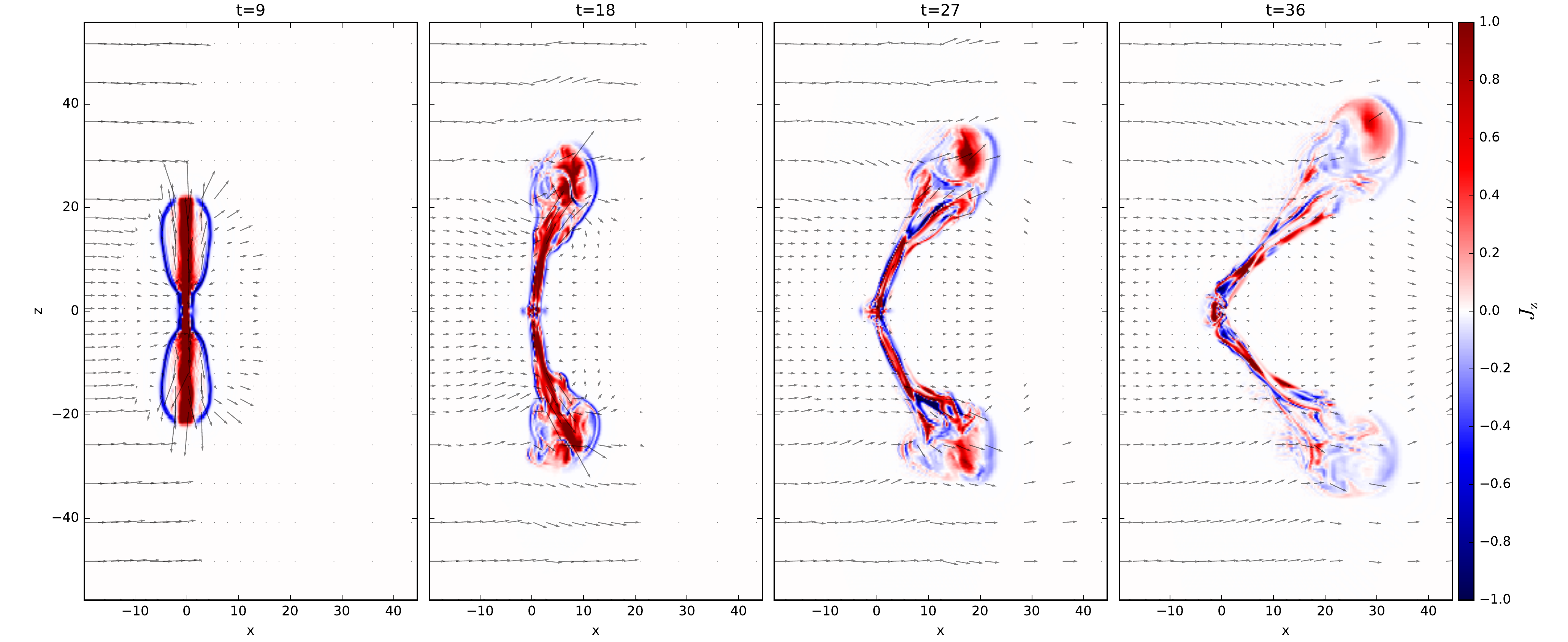}
\caption{Supplementary plots for Run W1. The pseudocolor maps (sliced from 3D data at $y=0$) in each panel, from top to bottom, show the pressure $p$, temperature $T$, normalized Bremsstrahlung emissivity ($j_{\rm brem} = \rho^2\sqrt{T}$) and the z-component of current $J_{\rm z}$, respectively. The plots in each column, from left to right, are the snapshots at $t=9,18,27,36$, respectively. The vectors in the bottom panel show the velocity field, and $T_0$ in the second panel (counted from top) is the temperature of the initial isothermal ICM background.}
\end{figure*}

In the upper panel of Figure \ref{fig:tower-jet-bending}, we show the bending process of a magnetic tower jet (Run W1). The pseudocolor map in each figure shows the density profile (sliced from 3D data at $y=0$), and the grey lines are the contours of magnetic intensity, which we use to indicate the jet morphology during the bending process.  The black-thick-solid line is the contact discontinuity (CD) surface between the wind and ICM (the wind-ICM CD surface, hereafter), and ahead the wind-ICM CD surface is the shock front driven by the wind (i.e. the ``wind shock" as shown in Run W0; note that the magnetic tower jet also drives another hydrodynamical shock by itself, i.e. the ``cocoon'', as shown in Run C1).  In Figure \ref{fig:tower-jet-bending-suppl}, we present some supplementary plots of the pressure $p$, temperature $T$, normalized Bremsstrahlung emissivity ($j_{\rm brem} = \rho^2\sqrt{T}$) and the z-component of current $J_{\rm z}$, beside density, for Run W1.

Although the jet bending is driven by the intra-cluster wind, we can see from the figures that there is no direct contact between the jet and wind during the whole bending process. Instead, it is in an {\it indirect} way that the wind affects the jet morphology, by modulating the intra-cluster environment through which the jet actually propagates \citep{jones_hot_1979}. Characterised by the passage of the wind shock and wind-ICM CD surface, the jet bending process could be divided into two phases:

\begin{enumerate}
\item
Jet impacted by the wind shock.
We can see from the upper panel of Figure \ref{fig:tower-jet-bending} that the bending process begins when the wind shock impacts the jet lobes. Since the jet is magnetically dominated, the magneto-sonic speed within the jet is much larger than its ambient fluid velocity $v_{\rm x}$. As a result, we can see that the shape of the lobes is not changed much during the bending process when compared to the ones in Run C1.  It is shown, in the leftmost two columns, that the wind shock passes through the jet quickly, leaving the jet only bent slightly after the passage of the wind shock.  

\item
Jet bent indirectly by the wind ram pressure.
After the passage of the wind shock, the jet is bent into a head-tail morphology following the compressive deformation of the downstream ICM ahead the wind-ICM CD surface. We can clearly see, from the rightmost two columns of Figure \ref{fig:tower-jet-bending}-\ref{fig:tower-jet-bending-suppl}, that the curvature of the jet morphology matches very well with the wind-ICM CD surface (the black-thick-solid lines). When comparing with Run W0 (of which the model setup is exactly the same as Run W1 but without a jet),  we can see that the jet could not modify significantly the bulk flow of the wind-ICM interaction, instead, the jet is forced to follow the ICM deforming process, while the later is driven by the wind ram pressure. For further confirmation, we compare the wind-ICM CD surfaces with those in Run W0 (the lower panel of Figure \ref{fig:tower-jet-bending}, dashed lines). The dashed lines in the upper panel of Figure \ref{fig:tower-jet-bending} are taken from the lower panel for the convenience of comparison. We can see that the wind-ICM CD surfaces almost superimpose completely with each other for the cases with and without a jet.
\end{enumerate}

It is important that the jet curvature matches the wind-ICM CD surface: 
i) since the dynamics of the wind-ICM interaction is almost independent on the jet dynamics as we discuss above, it could provide a way to study the intra-cluster environment of the head-tail sources by observing the jet curvature;  
ii) it implies that the jet bending process is actually determined by the intra-cluster environment (see, e.g., \citealp{jones_hot_1979}), rather than the pressure balance between the jet and wind (see, e.g., \citealp{begelman_twin-jet_1979}). In the analysis by Begelman et al., it is assumed that the jet engine is consistently active, and there is no ICM atmosphere around the jet engine. While in our simulation we try to mimic a realistic scenario for both the intra-cluster environment and the jet engine (e.g., with reasonable power and lifetime), so the jet bending process actually occurs after the jet engine already turns off --- the jet bending process in Run W1 takes $\Delta t \ge 27$ (164.7 Myr), which is much longer than a typical lifetime for AGN/jet engine (31 Myr in our simulations) as we argue at the beginning of this section.

It is worth noting that the magnetic tower jet could keep its integrity during the bending process. As shown in Figure \ref{fig:tower-jet-bending}-\ref{fig:tower-jet-bending-suppl}, we can see that the morphology of ``head-tail" structure can be well reproduced --- (1) two fluffy tails bent into the same direction; (2) no radio emission between the tails (since there is no magnetic field there); (3) X-ray cavities backfilled by radio emission, and also (4) core sloshing shown in X-ray images (e.g., enlongated X-ray halo as seen surrounding 3C 465).


On the contrary, as we demonstrate in \S\ref{sec:static-jet}, hydrodynamical instabilities could be easily excited in the pure hydrodynamical jet models, and consequently destroys the jet-lobe morphology. For the purpose of comparison, we also perform simulations on the jet bending processes of pure hydrodynamical jets. The model setup for the hydrodynamical runs is exactly the same as in Run W1, except that we replace the magnetic tower jet with the pure hydrodynamical ones. 

\begin{figure*}[t]
\label{fig:subsonic-jet-bending}
\centering
\includegraphics[width=0.75\textwidth]{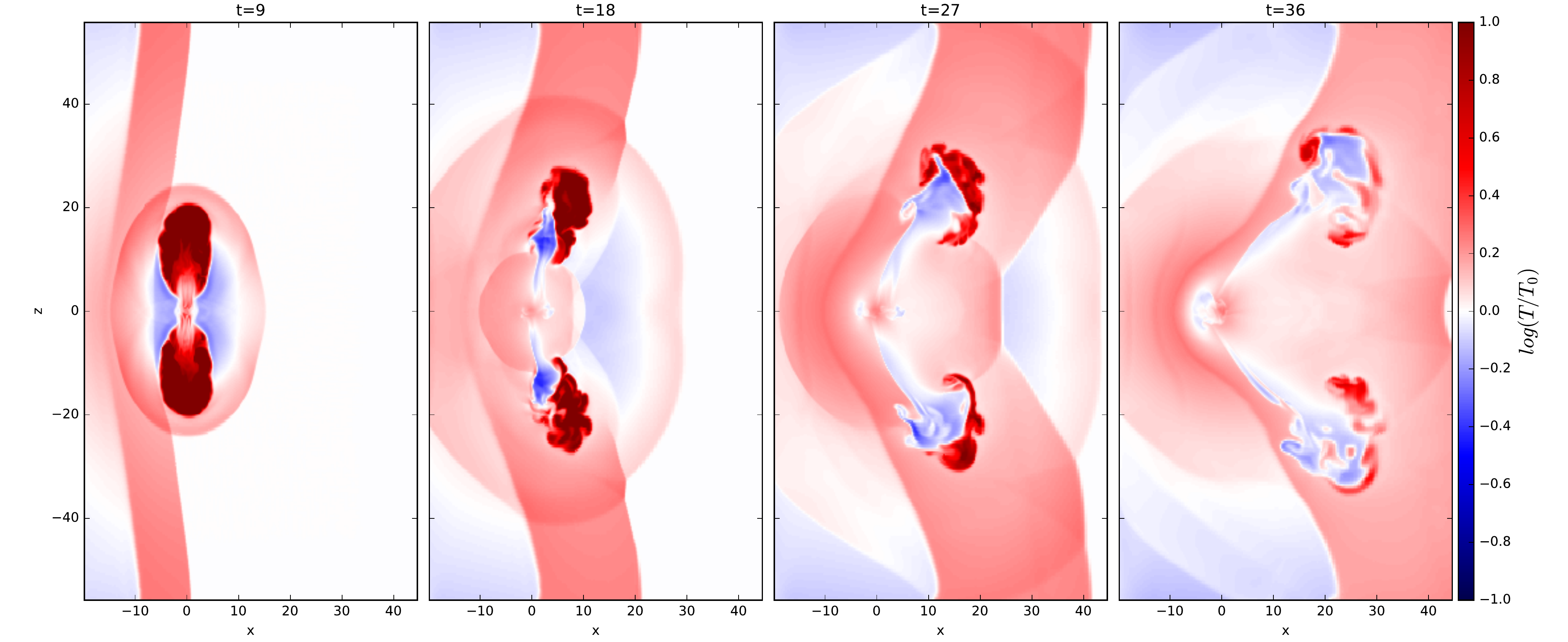}
\includegraphics[width=0.75\textwidth]{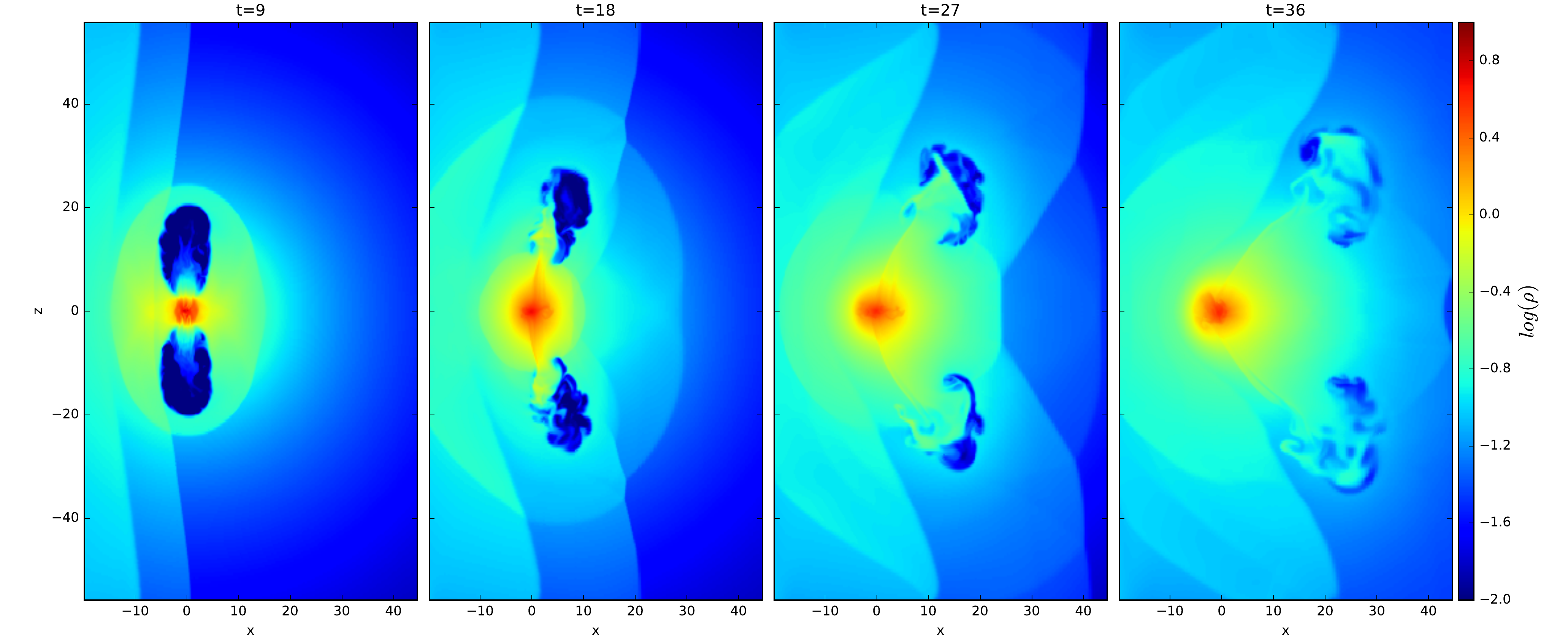}
\includegraphics[width=0.75\textwidth]{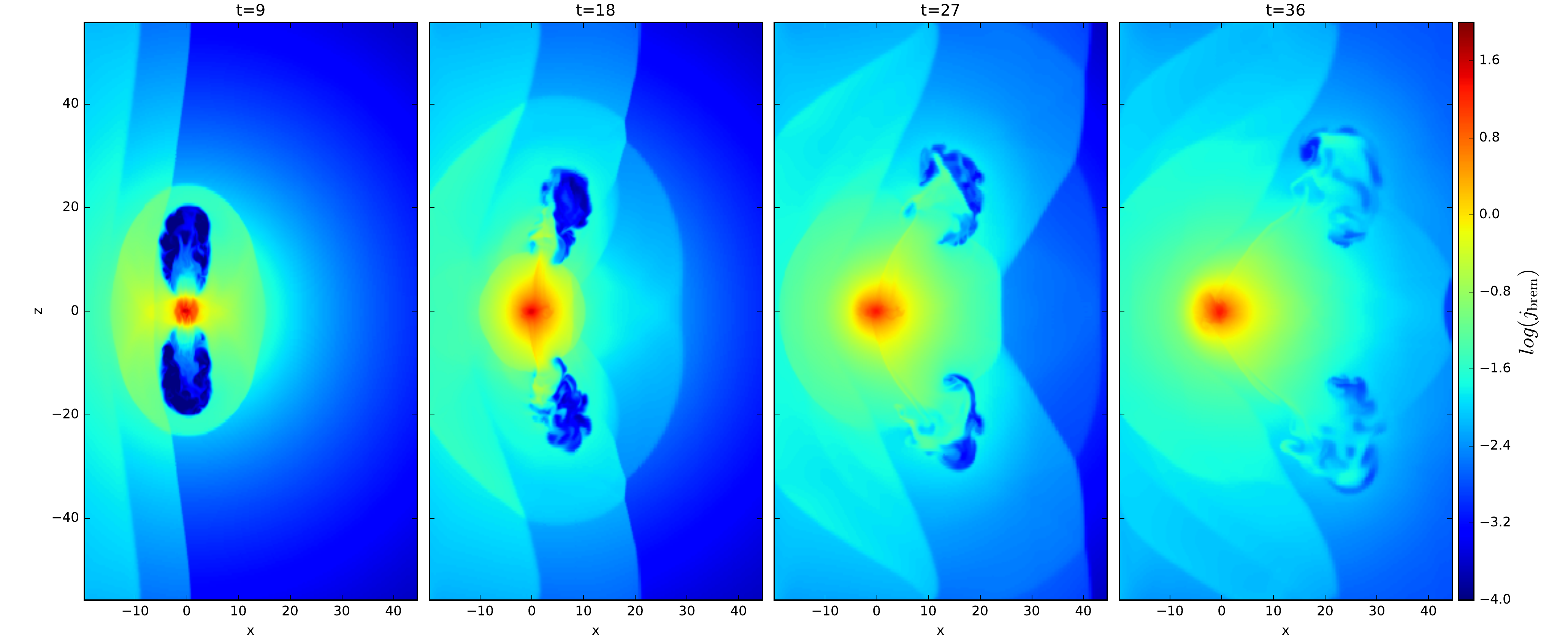}
\caption{Bending process of a subsonic hydrodynamical jet (Run W2). The pseudocolor map (sliced from 3D data at $y=0$) in each panel, from top to bottom, shows the profile of temperature, density, and normalized Bremsstrahlung emissivity ($j_{\rm brem} = \rho^2\sqrt{T}$), respectively. The plots in each column, from left to right, are the snapshots at $t=9,18,27,36$, respectively.}\end{figure*}
\begin{figure*}[t]
\label{fig:subsonic-jet-bending-suppl}
\centering
\includegraphics[width=0.75\textwidth]{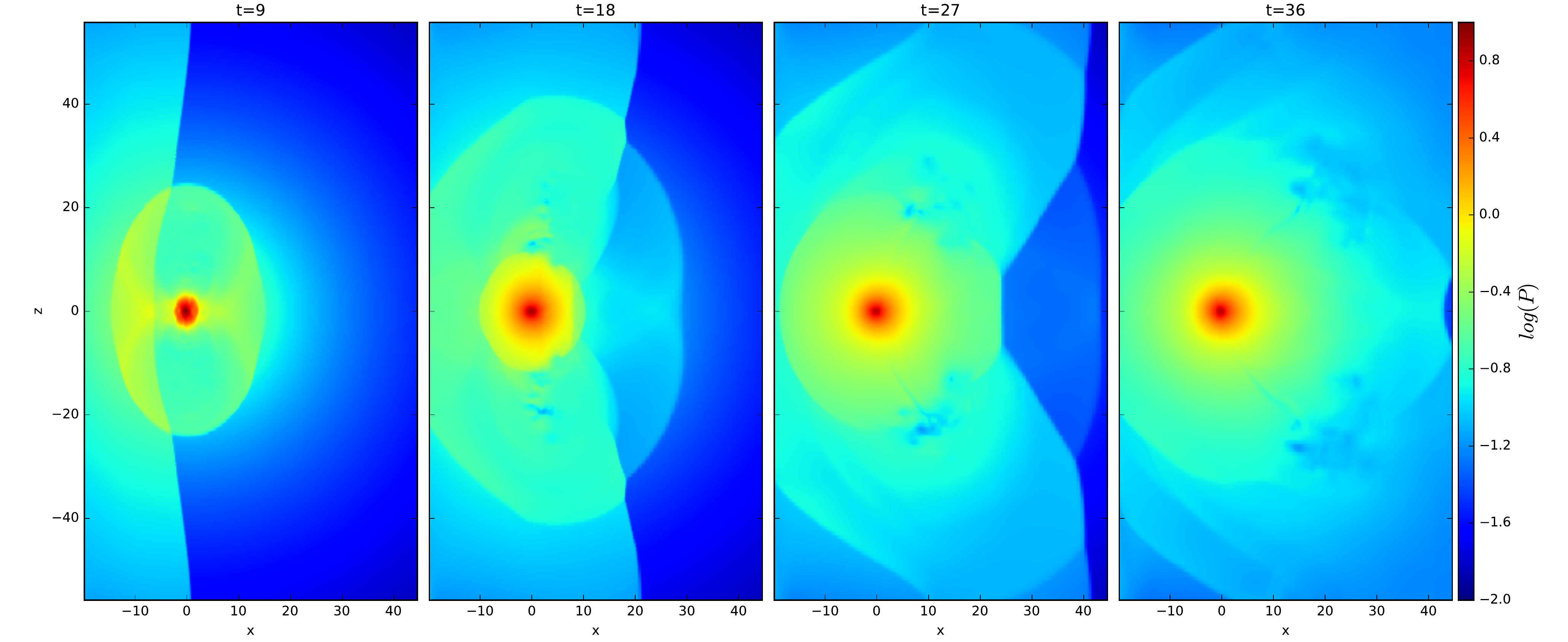}
\includegraphics[width=0.75\textwidth]{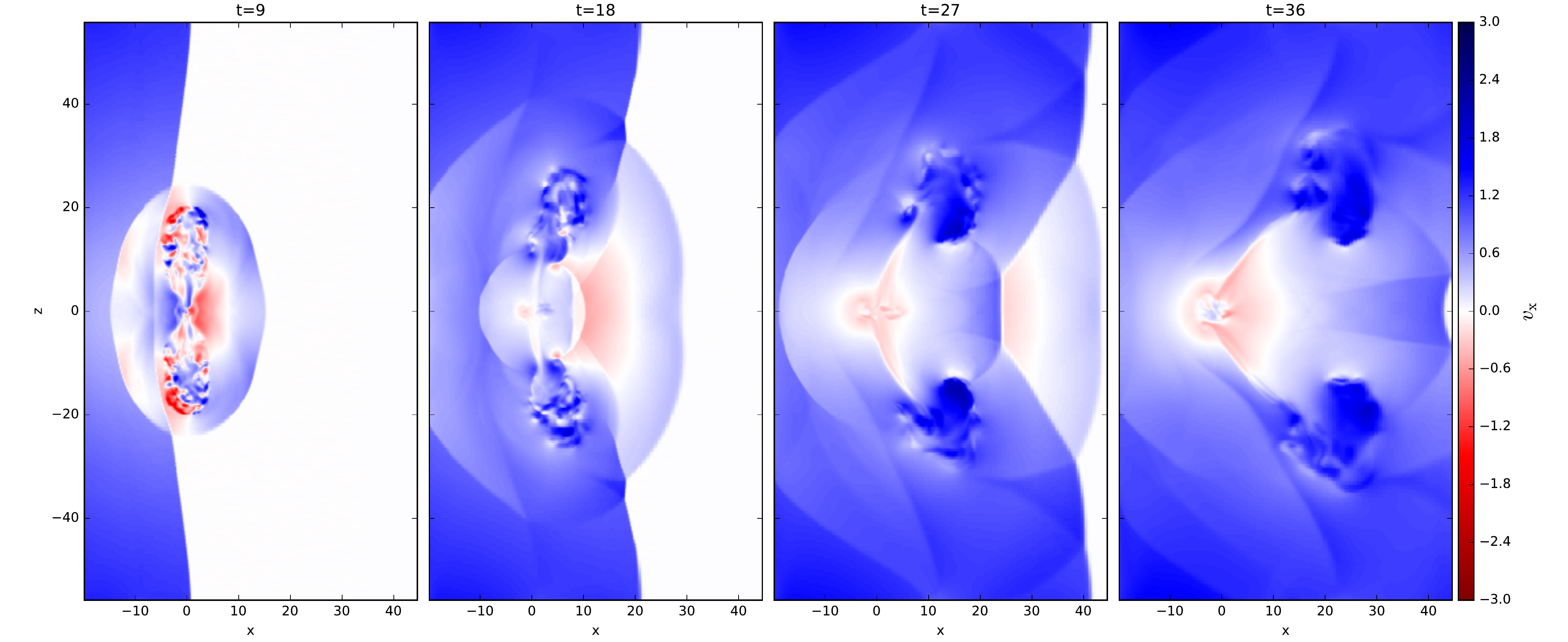}
\includegraphics[width=0.75\textwidth]{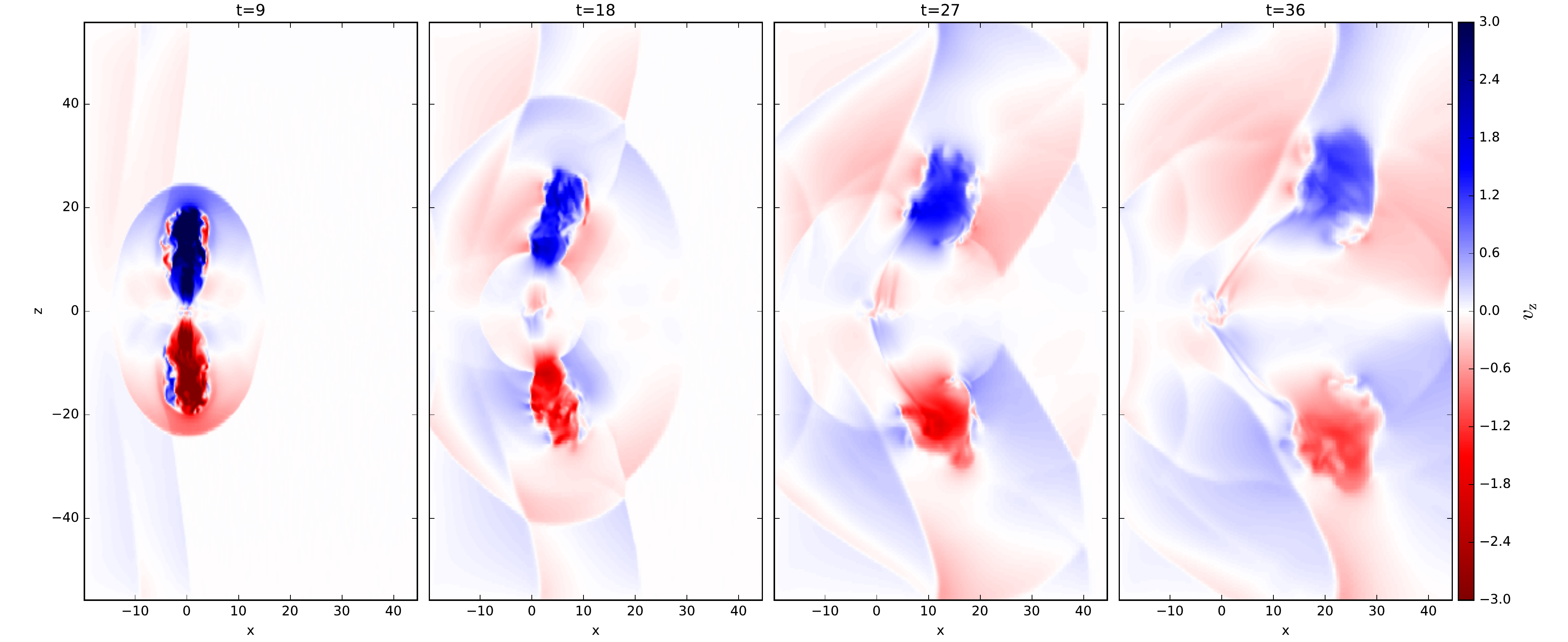}
\caption{Supplementary plots for Run W2. The pseudocolor maps (sliced from 3D data at $y=0$) in each panel, from top to bottom, shows the profile of the thermal pressure, the x- and z- component of velocity, respectively.}
\end{figure*}

In Figure \ref{fig:subsonic-jet-bending}-\ref{fig:subsonic-jet-bending-suppl}, we present the simulation results for the bending process of a subsonic hydrodynamical jet under the disruption of intra-cluster wind (Run W2).  The pseudocolor map in each panel shows the profile of temperature, density, and normalized Bremsstrahlung emissivity, etc. The plots in each column, from left to right, are the snapshots at $t=9,18,27,36$, respectively.
Similar to magnetic tower jet, the subsonic jet is also over-pressured (resulting in that the internal sound speed is much larger than the ambient fluid velocity), so that the jet lobes could quickly respond to the oncoming disruption, and be bent without  much compressive deformation. However, the subsonic jet is born to be unstable against the Rayleigh-Taylor instability as shown in Run C2, and in Run W2 the disruption by the wind makes the situation even worse. Finally, it turns out that the jet lobes could not keep their integrity during the bending process (see, e.g.,  the results at $t=36$). 

\begin{figure*}[t]
\label{fig:supersonic-jet-bending}
\centering
\includegraphics[width=0.75\textwidth]{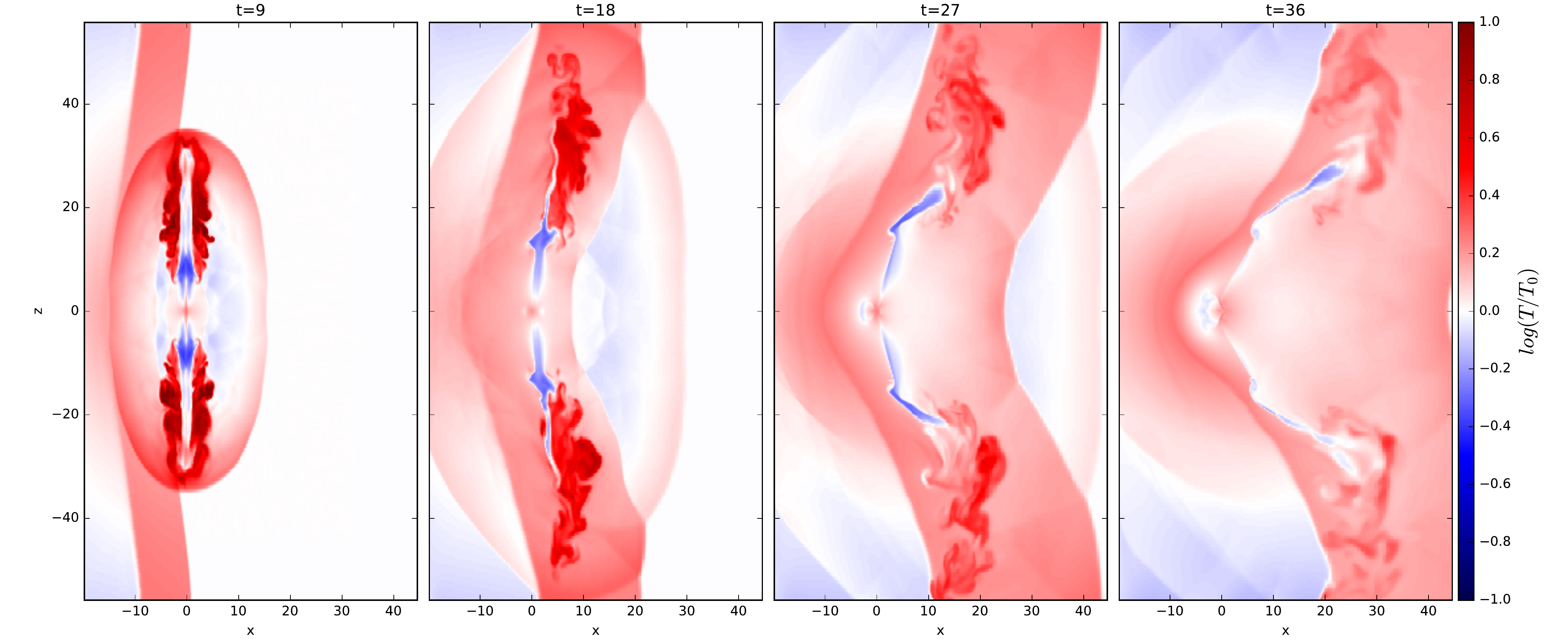}
\includegraphics[width=0.75\textwidth]{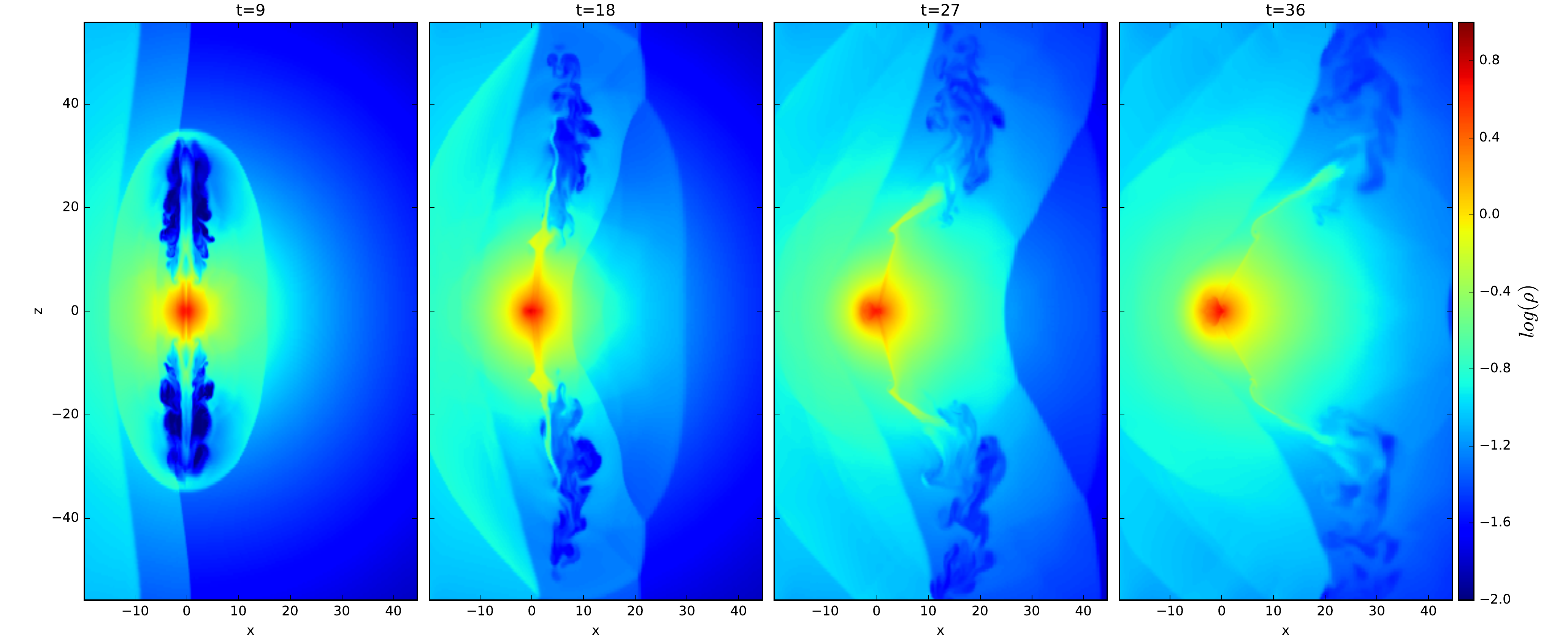}
\includegraphics[width=0.75\textwidth]{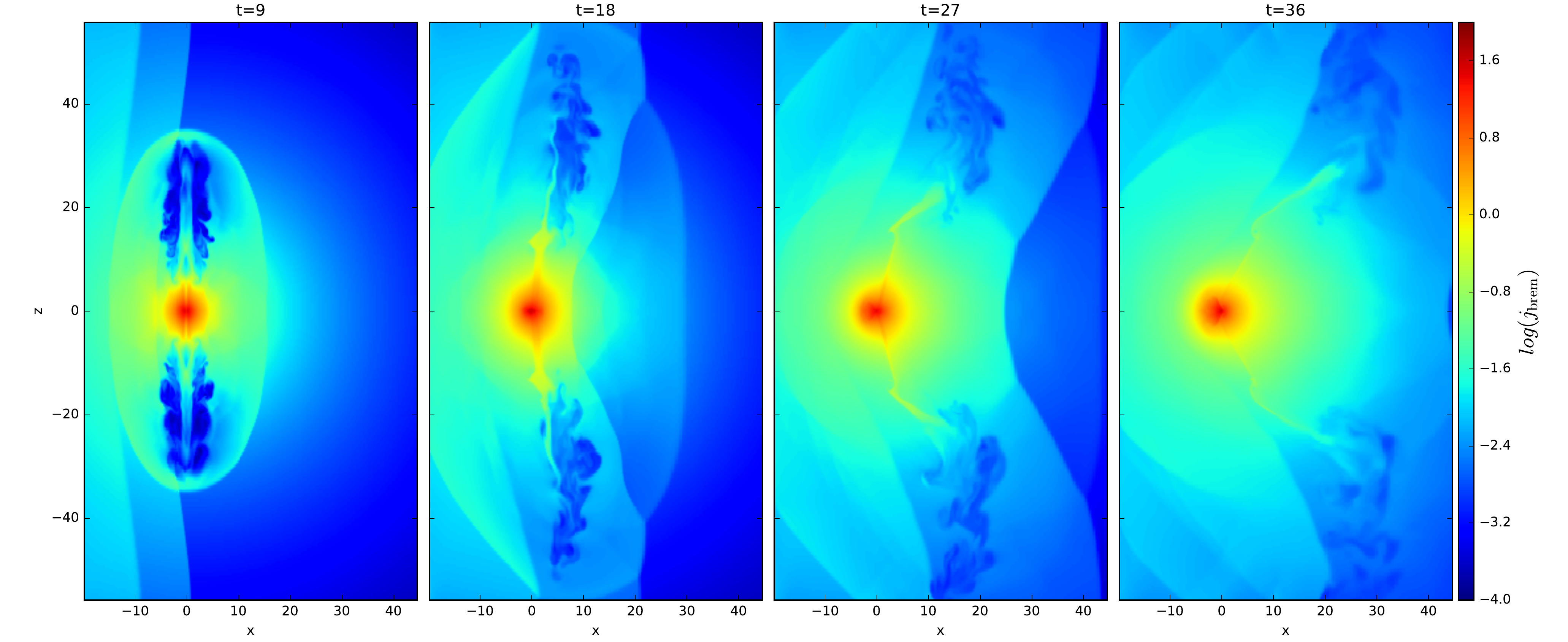}
\caption{Bending process of a supersonic hydrodynamical jet (Run W3). The pseudocolor map (sliced from 3D data at $y=0$) in each panel, from top to bottom, shows the profile of temperature, density, and normalized Bremsstrahlung emissivity ($j_{\rm brem} = \rho^2\sqrt{T}$), respectively. The plots in each column, from left to right, are the snapshots at $t=9,18,27,36$, respectively.}\end{figure*}
\begin{figure*}[t]
\label{fig:supersonic-jet-bending-suppl}
\centering
\includegraphics[width=0.75\textwidth]{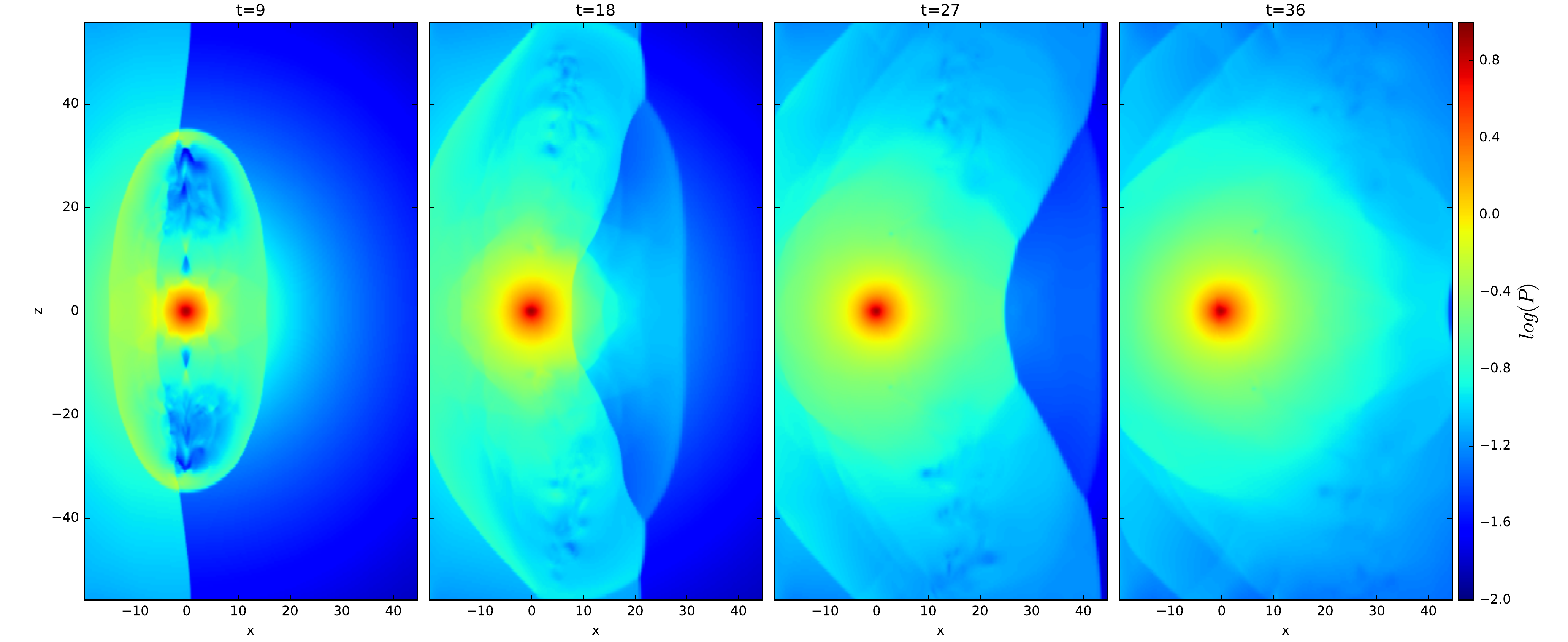}
\includegraphics[width=0.75\textwidth]{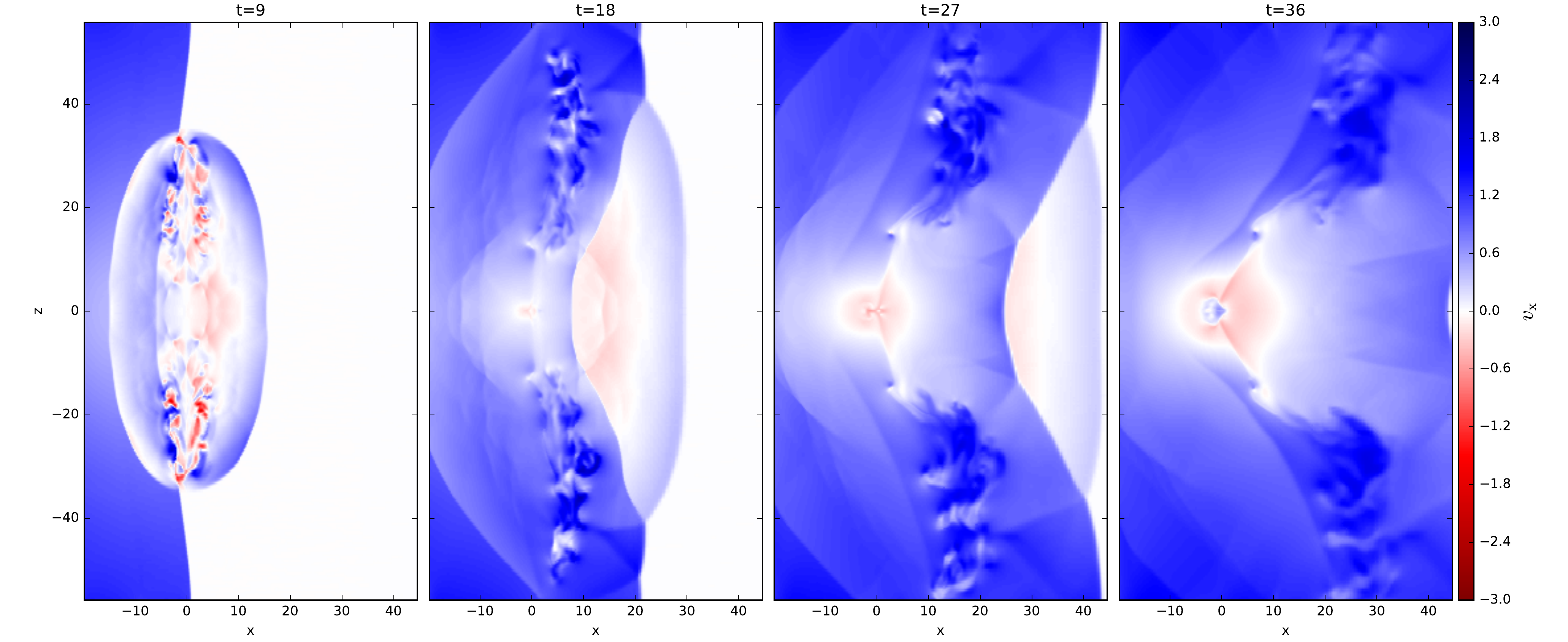}
\includegraphics[width=0.75\textwidth]{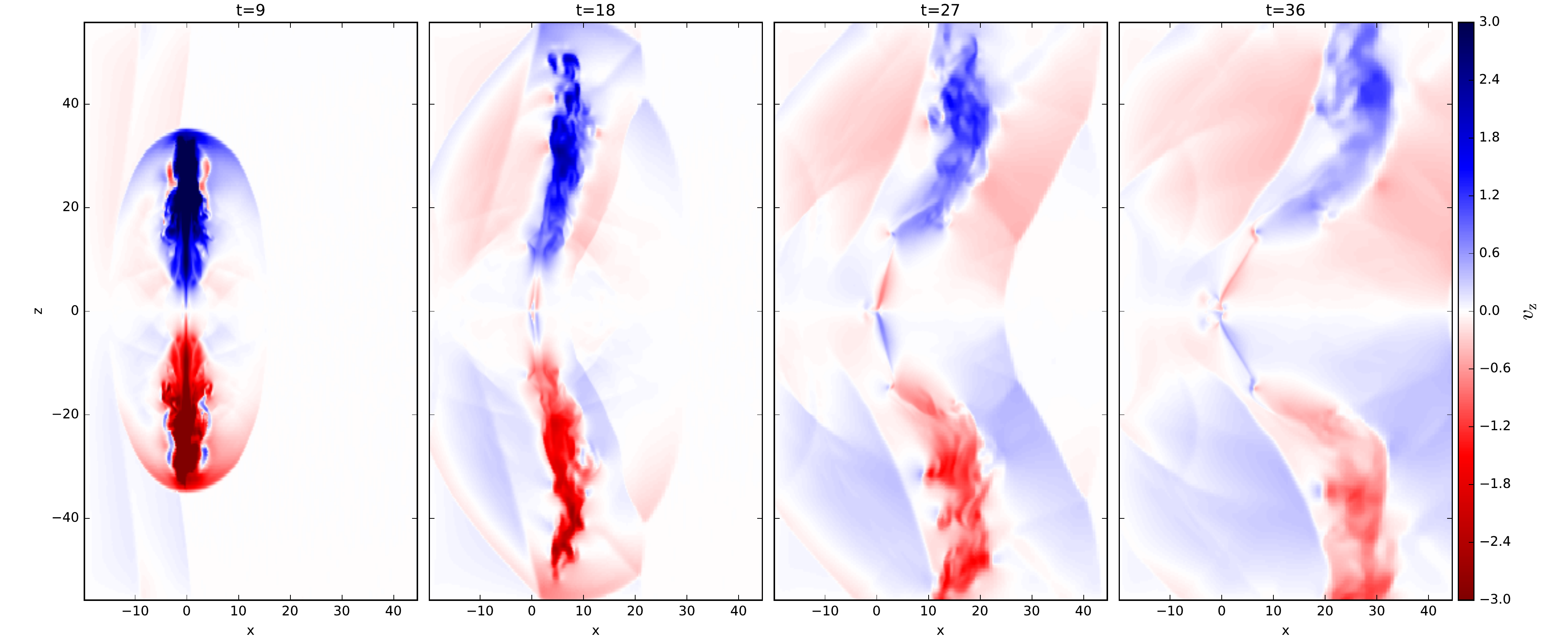}
\caption{Supplementary plots for Run W3.The pseudocolor maps (sliced from 3D data at $y=0$) in each panel, from top to bottom, shows the profile of the thermal pressure, the x- and z- component of velocity, respectively.}
\end{figure*}

In Figure \ref{fig:supersonic-jet-bending}-\ref{fig:supersonic-jet-bending-suppl}, we present the simulation results for the bending process of a supersonic hydrodynamical jet under the disruption of intra-cluster wind (Run W3). Compared to the subsonic hydrodynamical jet (Run W2), the velocity of the supersonic hydrodynamical jet is much larger,  consequently, there is even less time for the jet to form lobes, and the Kelvin-Helmholtz instability is excited and grows up quickly before/during the bending process. In addition, the supersonic jet is much colder (smaller sound sound) than the subsonic jet, we can see clearly some compressive deformation of the jet morphology during the bending process of the supersonic hydrodynamical jet. Like the case of subsonic hydrodynamical jet, the instabilities are so strong that the supersonic hydrodynamical jet is destructed rapidly when bent by the intra-cluster wind (see, e.g.,  the results at $t=36$).

For a brief summary of this subsection, in which we perform detailed analysis on the interaction among the magnetic tower jet, ICM and wind, we find that the jet bending process starts with the passage of the wind shock, and finally the jet is bent into head-tail morphology by the pressure gradient created by the wind-ICM interaction. 

Note that only the magnetic tower jet could survive from the bending process driven by wind, while the subsonic/supersonic hydrodynamical jet suffers severe hydrodynamical instabilities and could not survive under the strong disruption. 
Similar to the argument in \S\ref{sec:static-jet}, the reasons for the survival of magnetic tower jet are: (1) the plasma frozen-in effect and global current structure guarantee the integrity of the jet morphology; (2) magnetic fields could help to suppress the hydrodynamical instabilities; (3) the high internal magneto-sonic speed prevents the magnetic tower jet from significant compressive deformation during its bending process.

\subsubsection{Jet Bent by Shock}
\begin{figure*}[t]
\label{fig:jet-bent-by-shock}
\centering
\includegraphics[width=0.9\textwidth]{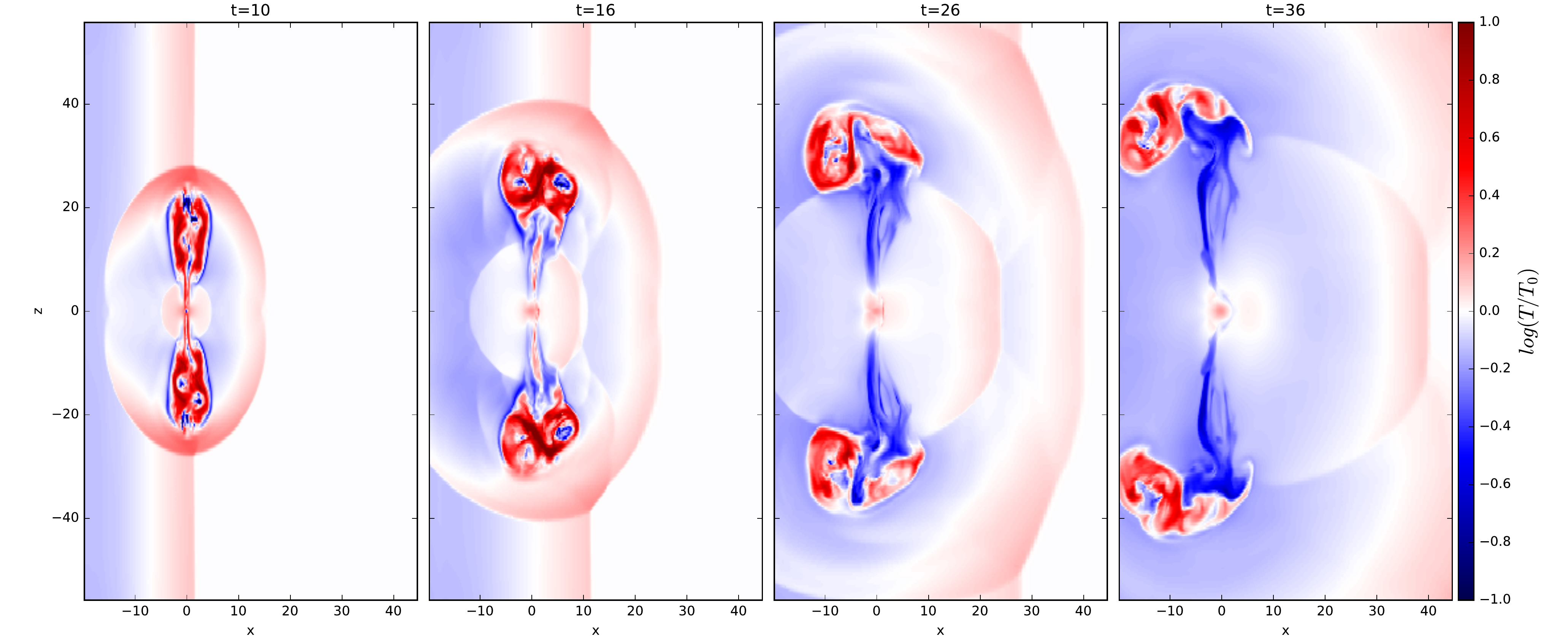}
\includegraphics[width=0.9\textwidth]{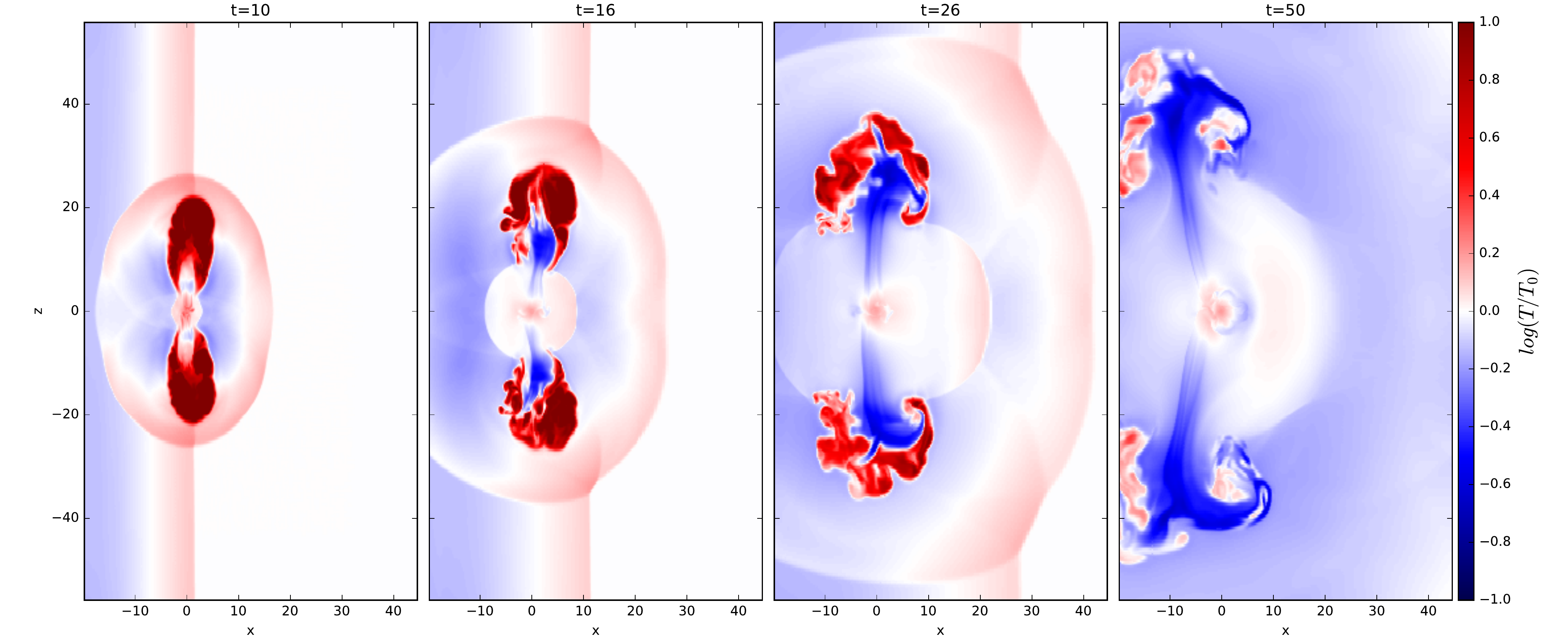}
\includegraphics[width=0.9\textwidth]{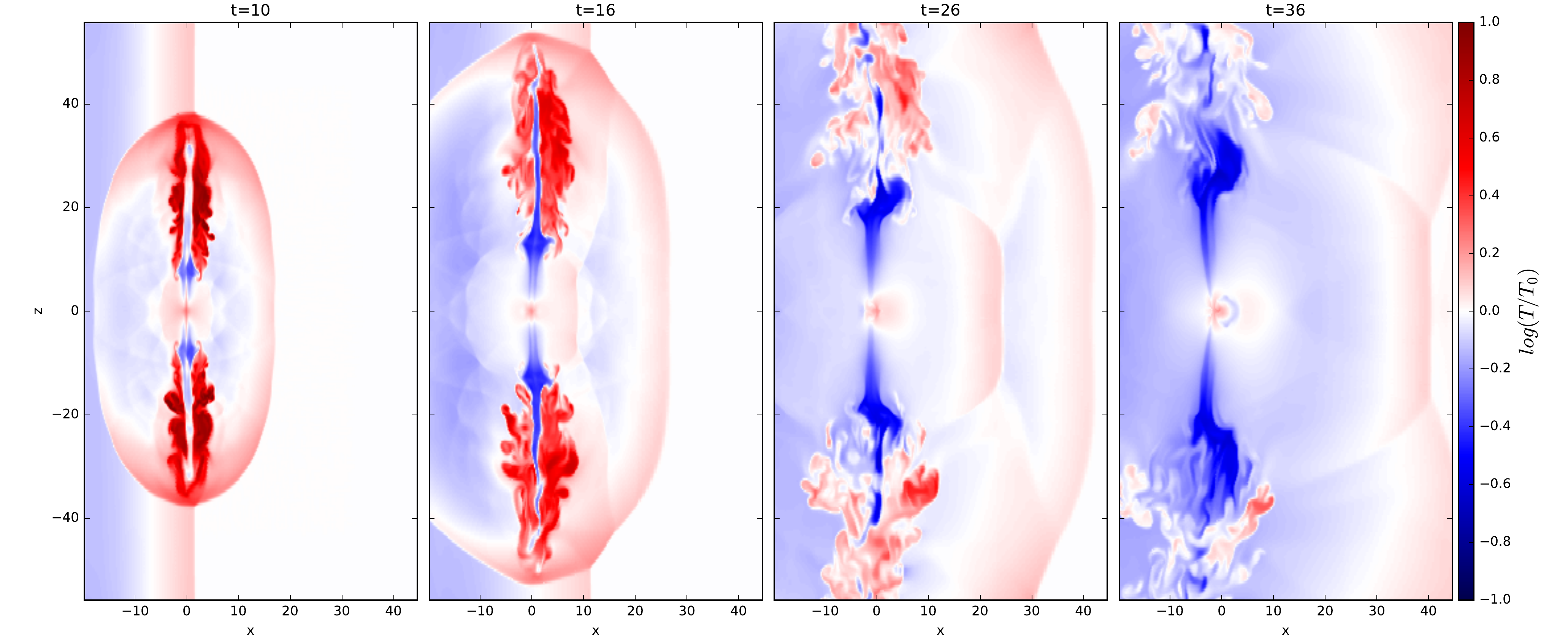}
\caption{Jet bent by shock only. The pseudocolor maps (sliced from 3D data at $y=0$) show the temperature profile in a time sequence, from left to right, with $t=10,16,26,36$, respectively. Each panel of the figures, from top to bottom, is for the case of magnetic tower jet (Run S1), subsonic hydrodynamical jet (Run S2), and supersonic hydrodynamical jet (Run S3), respectively. It is shown that the jets could be bent during the passage of the shock. However, the rarefaction wave behind the shock creates low pressure zone on the upstream of the jets, sucks the jets back and bends them into the opposite direction. }
\end{figure*}

As shown in \S\ref{sec:jet-wind-interaction}, shock is of the capacity to bend jet. In this subsection, we exam the possibility for jet bending by shock only. The model setup is the same as in \S\ref{sec:jet-wind-interaction}, except that we replace the wind with shock (see \S\ref{sec:intra-cluster-weather} for the detailed numerical implementation of shock). The simulation results are presented in Figure \ref{fig:jet-bent-by-shock}. The pseudocolor maps show the temperature profiles in a time sequence, from left to right, with $t=10,16,26,36$, respectively. Each panel of the figure, from top to bottom, is for the case of magnetic tower jet (Run S1), subsonic hydrodynamical jet (Run S2), and supersonic hydrodynamical jet (Run S3), respectively. 


It is shown that, during the passage of the shock, the jet could be bent to a magnitude ($\Delta x$) comparable to the lobe size. However, the rarefaction wave behind the shock succedently creates low pressure zone upstream from the shock, which tends to suck the jets back and bend them to the totally opposite direction. It turns out that the jet actually {\it wiggles} around the z-axis after the shock passage.  

\begin{figure}[t]
\label{fig:wind-shock-comparison}
\centering
\includegraphics[width=0.45\textwidth]{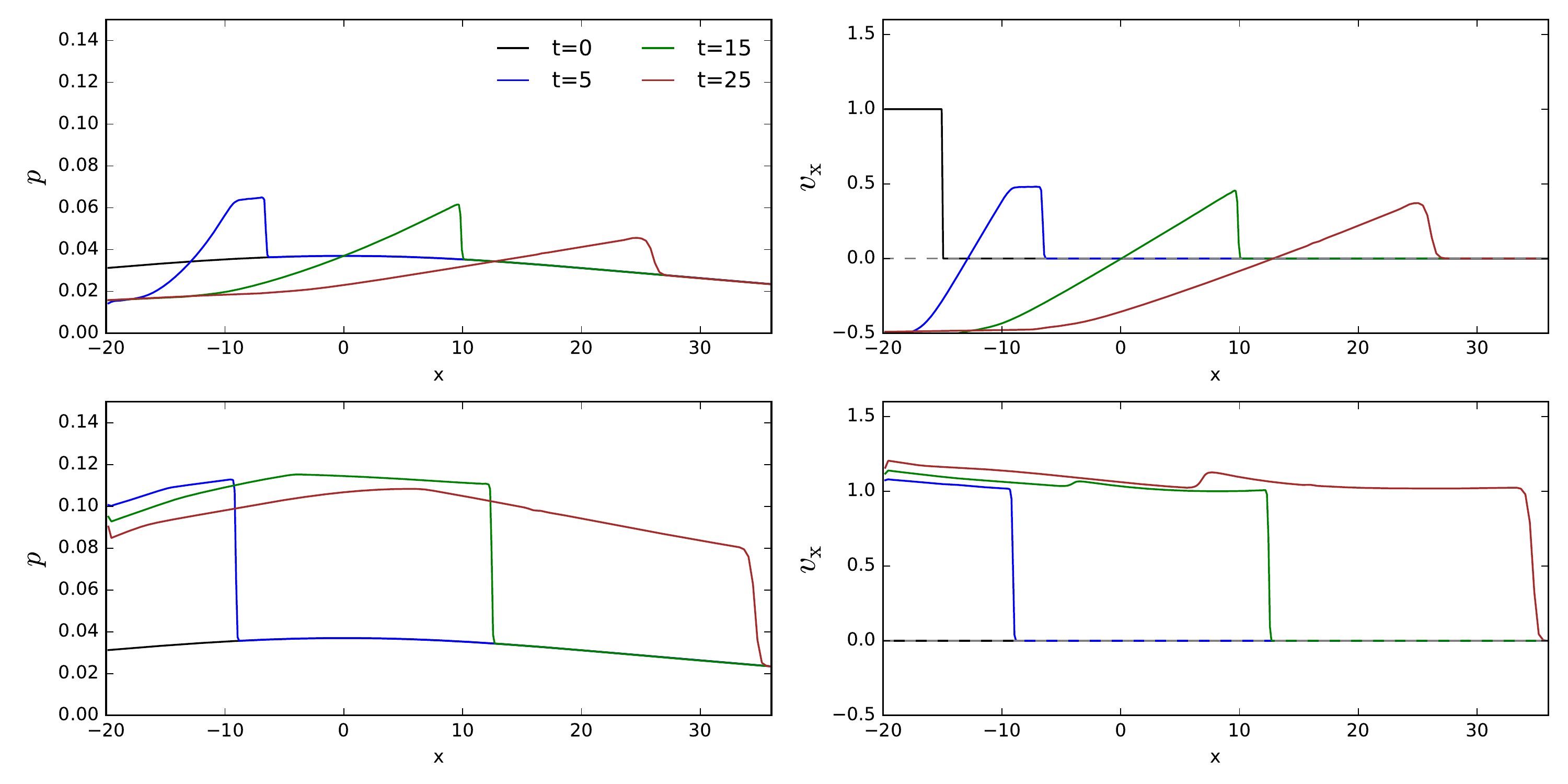}
\caption{Wind-shock comparison. The left panel shows the pressure profile in the x-direction (at $y=0, z=40$), while the right panel shows the profile of the x-component of fluid velocity. The upper panel is for the case with shock only (Run S0), and the results of the wind case (Run W0) is shown in the lower panel for the purpose of comparison.}
\end{figure}

To clarify the shock effects, we also present a simulation to study the interaction between the shock and the intra-cluster environment (Run S0), in which the model setup is the same as in Run S1 but without a jet. The results are presented in Figure \ref{fig:wind-shock-comparison}.  The left panel shows the pressure profile in the x-direction (at $y=0, z=40$), while the right panel shows the profile of the x-component of fluid velocity $v_{\rm x}$. The upper panel is for the case with shock only (Run S0), and the results of the wind case (Run W0) is shown in the lower panel for the purpose of comparison. It is shown that the ICM could gain some speed (positive $v_x$) only during the passage of the shock. Immediately after the shock passage, a low pressure zone is left behind (the rarefaction wave), and the ICM is forced to move backward by the pressure gradient (as a result, $v_{\rm x}$ decreases, and finally changes its sign). In contrast, in the wind case (W0), $v_{\rm x}$ keeps positive in the upstream, as momentum is continuously injected into the ICM (the lower-right panel). 

We conclude that shock could not bend the jet significantly, so could not be expected to explain the observed long tails in head-tail radio galaxies. Again, as shown in the right panel of Figure \ref{fig:jet-bent-by-shock}, the hydrodynamical jets (both the subsonic and supersonic ones) are destructed quickly by the  shock and the consequent rarefaction wave. While the magnetic tower jet could survive successfully from the violent disruption under the protection of its internal magnetic fields.

\section{Conclusion and Discussion} \label{sec:conclusion}

In this paper, we perform detailed three-dimensional magnetohydrodynamical simulations to study the behaviour of radio jet under violent intra-cluster weather.  We demonstrate that the distinctive morphology of the observed head-tail radio sources could be used to study the jet properties and also the ICM dynamics.

Three kinds of jet models are proposed, including the magnetic tower jet, subsonic hydrodynamical jet and supersonic hydrodynamical jet, and two kinds of intra-cluster weather are studied in this paper, i.e. wind (relative motion between the jet's host galaxy and its surrounding ICM) and shock  (e.g. induced by galaxy/cluster merger). To isolate the physical effect between the jet, ICM and wind/shock, we also study the wind-ICM interaction (without including a jet) and jet-ICM interaction (without intra-cluster weather), respectively.

\ 

The main conclusions are as follows:
\begin{itemize}
\item 
Core sloshing, as a result of the competition between the wind ram pressure and cluster gravity, is inevitable during the jet bending process. The sloshing pattern is asynchronous for different radii as the intrinsic dynamical timescale is a function of radius. Consequently, the asynchronous core sloshing induces significant shear gradient in the direction perpendicular to the wind velocity. Finally, the Kelvin-Helmholtz instability is excited and turbulence eventually grows up in the core region. {On the one hand, the shear motion/turbulence could help to homogenise the jet energy.} On the other hand, the shear motion of the sloshing ICM is so strong that it could easily disrupt the jet morphology, it is a challenge for jet to survive when propagating through the sloshing core.

\item
The morphology of hydrodynamical jets {\it could not survive} under the violent intra-cluster weather in all of the simulation runs we've made. As is well known, hydrodynamical jet always suffers the Kelvin-Helmholtz and Rayleigh-Taylor instabilities. The hydrodynamical instabilities induce turbulent edges along the jet path and could destroy the integrity of the jet morphology, making it difficult to form lobes even within a static environment. The existence of the intra-cluster weather makes the situation even worse. It is argued that viscosity might help to suppress the hydrodynamical instabilities and to keep the integrity of jet lobes \citep{reynolds_buoyant_2005}. however it is beyond the scope of this paper, and we leave it to the future work. 

\item
With the magnetic tower jet and a simplified intra-cluster wind model, we reproduce successfully the jet-lobe morphology similar to that of the observed head-tailed radio galaxies. Thanks to the protection of the internal magnetic fields, hydrodynamical instabilities are naturally suppressed, the magnetic tower jet could usually survive against the disruption of violent intra-cluster weather. It is found that the jet curvature matches well with the contact discontinuity surface between the wind and ICM, which is determined mainly by the hydrodynamical equilibrium between the wind ram pressure and ICM pressure (together with the cluster gravity), rather than the pressure balance between the jet and wind. As we argue in \S\ref{sec:results} that the lifetime of jet engine is usually much shorter than the timescale needed to bend the jet, the jet engine is inactive at most of the time during the bending process.  The jet tends to slow down and forms lobes on the hundred-kilo parsec scale, where the ram pressure of the jet becomes relatively small when compared to the wind ram pressure, i.e., it could hardly shift the profile of the wind-ICM interaction. Instead, the jet bending process then mainly follows the bulk motion of its surrounding ICM driven by the wind. 

\item
We also check the possibility for jet bending by shock only. We find that shock could not bend the jet significantly, and could not be expected to explain the observed long tails in head-tail radio galaxies.

\end{itemize}

The fact that the jet curvature is fully determined by the ICM dynamics modulated by the wind ram pressure can provide a way to study the intra-cluster environment of head-tail radio sources by observing their jet curvature (as shown in Run W1, for example). We shall emphasise that our model setup is quite different from those in which jet is bent directly by wind within an uniform low density background \citep[e.g.,][]{begelman_twin-jet_1979, morsony_simulations_2013}. In this paper, we are trying to simulate the jet bending process in a realistic intra-cluster environment, so the wind-ICM-jet interaction involves the ICM dynamics itself \citep{jones_hot_1979}. Based on the observed jet curvature, it is possible to infer, for examples, the wind speed/density, the ICM/gravity profile of the jet's host cluster. In addition, core sloshing is also providing constraints on the ICM dynamics. 

\begin{figure}[t]
\label{fig:tower-jet-lobe}
\centering
\includegraphics[width=0.475\textwidth]{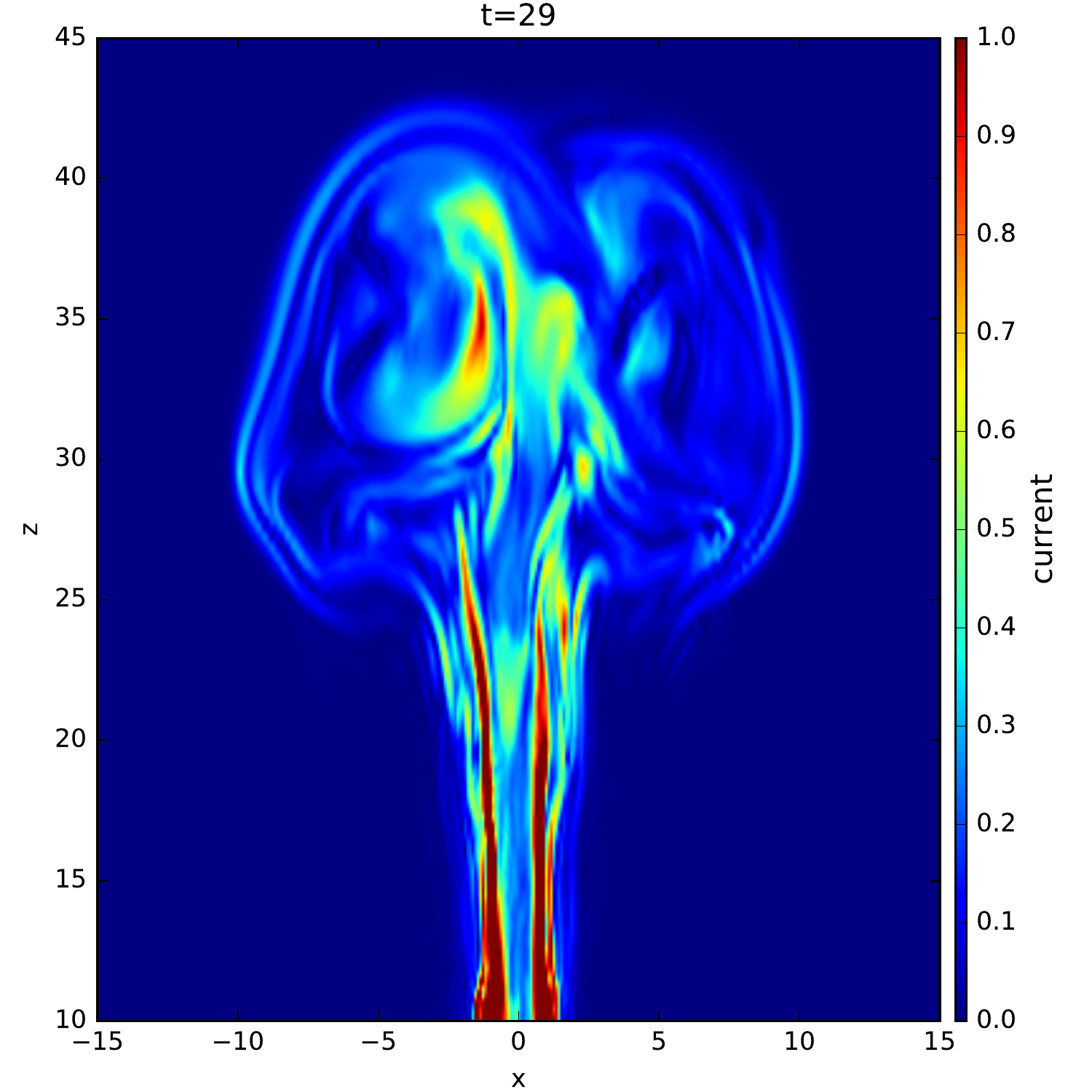}
\caption{Sub-structures in a lobe of the magnetic tower jet (Run C1). The pseudocolor map (sliced at $y=0$) shows the profile of current $|{\bf J}|$ at $t=29$. Current sheets and filamentary structures are found within the jet.}
\end{figure}

Furthermore, the jet-lobe morphology itself could also tell us a lot about the jet properties, e.g., (1) some compressive deformation of jet lobes is expected during the bending process. Theorectically, the larger internal (magneto-) sonic speed of the jet lobes, the smaller compressive deformation there should be during the bending process. As shown in  \S\ref{sec:jet-bending}, the compressive deformation of the jet lobes decreases in the order of supersonic hydrodynamical jet, subsonic hydrodynamical jet and magnetic tower jet; (2) magnetic tower jet is expected to have a sharp edge in its radio image as an effect of the return current. The internal kink instability could also induce some sub-structures  {\it inside the lobes} (see Figure \ref{fig:tower-jet-lobe}). In contrast, the instabilities in hydrodynamical jet models mainly occur {\it on the lobe surface}, which could also induce some distinguishable features on the jet-lobe morphology.

When we design the numerical experiments in this paper, we take into account the lifetime of the jet engine (i.e., fixed to 31 Myr, as a typical value of an AGN lifetime), which results in that the jet bending process occurs mainly when the jet engine already turns off. One should be aware that if the jet engine keep its activity during the bending process, the wind-ICM dynamics could be significantly altered, and so does the jet morphology, since huge amount of energy is put into the interaction.

As we argue above, core sloshing is inevitable in the presence of violent intra-cluster weather, and would induce turbulence in the late stage. To mimic the situation of 3C 465, we choose carefully the time occasion of jet-wind interaction so that it allows lobes to form at the early stage of core sloshing. Of course, it should be more common that the jet engine is active while the core already becomes sloshing. However, it is a challenge for the jet engine to survive against the strong shear motion in a sloshing core, moreover, our implementation of the jet engine is actually detached from the central supermassive black hole (where there should be no sloshing), so it is technically limited to simulate such situations. 
{Also, we note that the jet could have some rotation but our analyses show that the effect of rotation is small and it does not affect the stability of our jets (though see, e.g., \citealp{ciardi_curved_2008} on how jet rotation affects the jet stability). We leave those to the future work to improve the model setup.}

\acknowledgments{
HL acknowledge useful discussions with J. Eilek and F. Owen. HL \& SL acknowledge support by LANL's LDRD and DOE/CMSO support. 
ZG \& FY acknowledge support by the Natural Science Foundation of China (grants 11573051 and 11633006), the Key Research Program of Frontier Sciences of CAS (No.  QYZDJ-SSW-SYS008), the Strategic Priority Research Program ``The Emergence of Cosmological Structures'' of CAS (grant XDB09000000),  the grant from the Ministry of Science and Technology of China (No. 2016YFA0400704), and the CAS/SAFEA International Partnership Program for Creative Research Teams.}

\bibliography{reference-list}

\end{document}